\documentstyle[12pt,ssi,epsf]{article}
\begin{document}
\title{Gravitational radiation sources and signatures}
\author{Lee Samuel Finn\thanks{Supported by National Science
    Foundation awards PHY 98-00111 and PHY 95-03084 to The 
    Pennsylvania State University, and PHY 93-08728 to Northwestern 
    University.}\\
Center for Gravitational Physics and Geometry\thanks{Also 
Department of Physics, and Department of Astronomy and Astrophysics}, 
The Pennsylvania State University, University Park,
  Pennsylvania 16802}
\maketitle
%\begin{abstract}
%\end{abstract}
\section{Introduction}

The goal of these lecture notes is to introduce the developing
research area of {\em gravitational-wave phenomenology.} In more
concrete terms, they are meant to provide an overview of
gravitational-wave sources and an introduction to the interpretation
of real gravitational wave detector data.  They are, of course,
limited in both regards.  Either topic could be the subject of one or
more books, and certainly more than the few lectures possible in a
summer school.  Nevertheless, it is possible to talk about the
problems of data analysis and give something of their flavor, and do
the same for gravitational wave sources that might be observed in the
upcoming generation of sensitive detectors.  These notes are an
attempt to do just that.

Despite an 83 year history, our best theory explaining the workings of
gravity --- Einstein's theory of general relativity --- is relatively
untested compared to other physical theories.  This owes principally
to the fundamental weakness of the gravitational force: the precision
measurements required to test the theory were not possible when
Einstein first described it, or for many years thereafter.

The direct detection of gravitational-waves is a central component of
our first investigations into the dynamics of the weakest of the known
fundamental forces: gravity.  It is only in the last 35 years that
general relativity has been put to significant test.  Today, the first
effects of static relativistic gravity beyond those described by
Newton have been well-studied using precision measurements of the
motion of the planets, their satellites and the principal asteroids. 
Dynamical gravity has also been tested through detailed and
comprehensive observations of the slow, secular decay of the
Hulse-Taylor binary pulsar system \cite{taylor82a,taylor89a}.  What
has not heretofore been possible is the direct observation of the
effects of dynamical gravity on a laboratory instrument: {\em i.e.,}
the direct detection of gravitational radiation.

The scientific importance of the direct detection of
gravitational-waves does not stop with its detection, however.  Strong
gravitational-waves are difficult to generate: so difficult, in fact,
that there is no possibility of a gravitational-wave ``Hertz''-type
experiment, where both the source and receiver are under laboratory
control.  The strongest gravitational-waves incident on Earth, as
measured by our ability to detect them in the sensitive detectors now
under construction, arise from astronomical sources.  These are also
the only sources that we can hope to observe in our detectors.  The
strongest of these anticipated sources --- inspiraling or colliding
neutron stars or black holes --- are, in fact, of cosmological origin.

Very little relevant detail is known about the gravitational-wave
sources that we anticipate may be detectable in the instruments now
under construction.  Estimates of source strengths and event rates are
difficult to make reliably.  This is because, at a deep and
fundamental level, our understanding of the cosmos is limited to what
we can learn from photons.  The mechanism by which gravitational-waves
are generated, on the other hand, favors sources that either do not
radiate electromagnetically ({\em e.g.,} black holes), are obscured
from view ({\em e.g.,} the gravitational collapse of a stellar core),
or are so distant and decoupled from the immediate origin of the
corresponding electromagnetic radiation that we cannot reliably
decipher the relevant source characteristics from the photons that
reach us ({\em e.g.,} $\gamma$-ray bursts).\cite{mohanty99a}

Gravitational-wave observations thus add a new dimension to our
ability to observe the Universe: the observations that we make will
tell us things we don't already know through other means.

In order to describe sensibly the signature of gravitational-wave
sources in real detectors we must first discuss in some detail how we
characterize gravitational-waves, how we characterize
gravitational-wave detectors, and how we give operational meaning to
the word ``detect''.  These three subjects are addressed in sections
\S\ref{sec:charGR}, \S\ref{sec:charDet} and \S\ref{sec:charToDet},
respectively.  In the context of gravitational-wave detection,
gravitational-wave signals fall fairly neatly into three categories:
burst signals, periodic signals and stochastic signals.  Sources
thought to be responsible for detectable signals in these categories
are described in sections \S\ref{sec:bursts}, \S\ref{sec:periodic} and
\S\ref{sec:stochastic}, respectively.

\subsection{Conventions}\label{sec:conventions}

\begin{itemize}
  
  \item The distance from detector to source will always be large
  compared to either a wavelength of the radiation field or the
  physical dimension of the detector; consequently, the incident
  radiation is effectively planar.
  
  \item We choose a sign convention for the line element of Minkowskii
  spacetime and recall the {\em Einstein summation convention,}
  wherein repeated Greek indices in a product are implicitly summed
  over their full range:
  \begin{eqnarray}
    ds^{2} &=& \eta_{\mu\nu}\,dx^{\mu}\,dx^{\nu}\\
    &=& \sum_{\mu,\nu=0}^{3} \eta_{\mu\nu}\,dx^{\mu}\,dx^{\nu}\\
    &=& -c^2 \left(dx^0\right)^2 + \left(dx^1\right)^2 
    + \left(dx^2\right)^2 + \left(dx^3\right)^2\\
    &=& -c^{2}dt^{2}+dx^{2}+dy^{2}+dz^{2}.
  \end{eqnarray}
  
  \item We will always treat the gravitational fields as weak and use 
  coordinates on spacetime that are either ``Cartesian''$+$time or 
  ``spherical-polar''$+$time.
    
  \item We will find it convenient to introduce the use of Latin
  indices to represent just the spatial components ({\em i.e.,} $x$,
  $y$, and $z$) of a tenser.  We generalize the Einstein summation
  convention to apply to repeated Latin indices in a product
  expression, with the implicit sum running over just the spatial
  coordinates, {\em e.g.,}
  \begin{equation}
      x^iy_i = \sum_{j=1}^{3} x^jy_j.
  \end{equation}

\item Except in the first several sections of these lecture notes, we
  will always work in units where the Newtonian gravitational constant
  $G$ and the speed of light $c$ are numerically equal to unity and
  the appearance of these constants in various formulae will be
  suppressed.  Dimensional analysis will always suffice to determine
  how the formulae should appear with the constants in place.
    
  \item Using $c$ we can express time in units of length and frequency
  in units of inverse length; similarly, by exploiting $G$ and $c$ we
  can express mass and energy in units of length.  Power is then a
  dimensionless number.  For CGS units, the conversion factors between
  mass, energy and length, and the physical constant with units of
  power, are
  \begin{eqnarray}
    G/c^{2} &=& 7.42\times10^{-29} \,\mbox{cm/gm} \label{eq:units-mass}\\
    G/c^{4} &=& 8.26\times10^{-50} \,\mbox{cm/erg} \label{eq:units-energy}\\
    c^{5}/G &=& 3.63\times10^{59} \,\mbox{erg/s} \label{eq:units-power}.
  \end{eqnarray}
\end{itemize}
  
\section{Characterizing Gravitational Radiation}\label{sec:charGR}

For our purpose here --- recognizing gravitational waves incident on a
detector --- two different characterizations of gravitational
radiation are useful.  The first is the radiation waveform and the
second is the signal ``power spectrum''. The waveform describes the 
radiation field's time dependence while the power spectrum 
describes its Fourier components. In \S\ref{sec:radwave} and
\S\ref{sec:spectrum} we describe these different characterizations of
gravitational radiation.  Several important physical insights
regarding gravitational radiation sources can be gained by considering
the instantaneous power radiated by a source: we discuss these
insights in \S\ref{sec:insights}.

\subsection{Radiation waveform}\label{sec:radwave}

In this subsection we review briefly the expression of the radiation
incident on a detector. Much of this section is by way of review; for
more details, see either the lectures by Bob Wagoner in this
collection, one of the many text books on
relativity\cite{weinberg72a,misner73a,schutz90a,stephani93a,ohanian94a},
or an excellent review article on the subject.\cite{thorne80a}

Gravitation manifests itself as spacetime curvature and gravitational
waves as ripples in the curvature that appear to us, moving through
time, to be propagating.  Detectors are generally not directly
sensitive to curvature, but to the mechanical displacement of their
components; so, we focus our attention on the spacetime metric, from
which physical distances between points in spacetime are determined. 
(The curvature is a function of the metric's second derivatives.)

We assume that gravity is weak in and around our detector;
correspondingly, we treat the spacetime metric as if it were the
metric of Minkowskii spacetime, plus a small perturbation:
\begin{equation}
  g_{\mu\nu} = \eta_{\mu\nu} + h_{\mu\nu},
\end{equation}
where $\eta_{\mu\nu}$ is the Minkowskii metric and $h_{\mu\nu}$ the 
metric 
perturbation. The corresponding line element, describing the proper
distance between nearby spacetime events whose coordinate separation is the
infinitesimal $dx^\mu$, is
\begin{equation}
ds^2 = g_{\mu\nu}dx^\mu dx^\nu = 
\eta_{\mu\nu}dx^\mu dx^\nu + h_{\mu\nu}dx^{\mu}dx^{\nu}.
\end{equation}
Detecting gravitational waves amounts to building instruments that are
sensitive to the effects of the small perturbation $h_{\mu\nu}$;
determining the signature of the gravitational waves in the detector
thus requires determining $h_{\mu\nu}$ and its influence on the
detector.

The metric $g_{\mu\nu}$ tells us how the proper distance between
points in spacetime is associated with our choice of coordinate
system. Since the gravitational fields near our detector are weak
and the spacetime nearly Minkowskii, we can introduce coordinates
that are, in the neighborhood of the detector, nearly the usual
Minkowskii-Cartesian coordinates, with the deviations of the
order of the perturbation.

Now, small changes in the coordinate system do not change the proper
distance between events, only our labeling of them.  If we make small
changes in our coordinate system, of the order of the perturbation,
then we will make corresponding changes in the perturbation
$h_{\mu\nu}$.  We can use this freedom to simplify the expression of
$h_{\mu\nu}$.  Coordinate changes do not change the physics or any
observable constructed from $h_{\mu\nu}$, of course.  For this reason,
and in analogy with electromagnetism, coordinate choices like these
are referred to as {\em gauge\/} choices.

With the separation of the metric into the Minkowskii metric
$\eta_{\mu\nu}$ plus a small perturbation, the field equations of
general relativity become (at first order in the perturbation) a set
of second order, linear differential equations for the ten components
of the symmetric $h_{\mu\nu}$.  Consequently, fixing the coordinates
allows us to impose eight conditions on the ten components of the 
symmetric $h_{\mu\nu}$,
leaving just two dynamical degrees of freedom.  These are identified
as the two polarizations of the gravitational radiation field.

An important gauge choice, always possible for radiative perturbations
about Minkowskii space, is the {\em Transverse-Traceless,} or TT, gauge.
Transverse-Traceless gauge is always associated with a particular
observer of the radiation. Let the four-velocity of this observer have
components $U^{\mu}$.  Without loss of generality let $t$ mark the
proper-time of this observer (so that $U^\mu$ is just the coordinate
vector in the $t$ direction) and $x$, $y$ and $z$ be the usual Cartesian
coordinates (to ${\cal O}(h)$) in the neighborhood of the observer. 
In TT-gauge the field equations are 
\begin{eqnarray}
\left(-{\partial^{2}\over\partial t^{2}}+\nabla^{2}\right)h_{\mu\nu} = 
-16\pi T_{\mu\nu},
\label{eq:wave-eq}
\end{eqnarray}
subject to the constraints
\begin{eqnarray}
    h_{\mu\nu}U^\mu &=& 0,\\
    {\partial\over\partial x^{k}}\left(h_{j\mu}\eta^{\mu k}\right)
     &=& h_{j}{}^{k}{}_{,k} = 0,\\
    h_{jk}\delta^{jk} &=& 0;
\end{eqnarray}
The metric perturbation satisfies a wave equation whose source is the
{\em stress-energy density\/} $T_{\mu\nu}$ of the matter and
(non-gravitational) fields.  In more physical terms, the constraints
are (in order):
\begin{itemize}
    \item $h_{\mu\nu}$ is purely spatial;
      
    \item the (spatial) metric perturbation $h_{ij}$ is purely
      transverse: {\em i.e.,} if the radiation wavevector is $k^{i}$
      (where the index $i$ runs over just the spatial dimensions; see
      \S\ref{sec:conventions}), then $h_{ij}k^{i}$ vanishes for all
      $j$; and
    
    \item the metric perturbation is {\em trace-free.} 
\end{itemize}
When there might be confusion we denote a metric perturbation in
TT-gauge with a superscript TT on $h_{\mu\nu}$; also, since the
perturbation is purely spatial, we generally refer just to the spatial
projection (in the coordinate system of the observer) of the
perturbation, as in $h^{TT}_{ij}$.

Given a metric perturbation $h'_{\mu\nu}$, expressed in any gauge,
corresponding to a plane wave propagating in the direction $n^k$, we
can recover the corresponding metric perturbation in TT-gauge by
applying the linear operator $P_{lm}$ to the spatial $h'_{ij}$:
\begin{eqnarray}
    h^{TT}_{ij} &=& 
    P_{i}^{l}h'_{lm}P^{m}_{j}-{1\over2}P_{ij}P^{lm}h'_{lm}\\
    P_{lm} &=& \delta_{lm} - n_{l}n_{m}.
\end{eqnarray}
Here and henceforth we will always express the metric perturbation in
TT-gauge. 

As mentioned above, gravitational wave detectors work by sensing the
relative motion of their components induced by a passing gravitational
wave.  Let's see how the TT-gauge metric perturbation is related to
such relative motion.

Consider a single isolated test mass, initially at rest at coordinate
position $\vec{x}_{A}$ in a TT-gauge coordinate system.  No forces act
on this test mass; so, it moves through spacetime in such a way that
its four-velocity always remains tangent to itself.  (Forces, of
course, cause the four-velocity to change direction.)  The
corresponding equations of motion for the spatial coordinates of the
test mass are
\begin{equation}
    {d\over d\tau}\left(dx^{i}_{A}\over d\tau\right) +
    \Gamma^{i}_{\mu\nu}{dx^{\mu}_{A}\over d\tau}{dx^{\nu}_A\over d\tau} = 0,
\end{equation}
where $\tau$ is the proper time of the test mass (initially $\tau$ is
equal to $t$ since the test mass is at rest) and
$\Gamma^{\alpha}_{\beta\gamma}$ is the metric connection
\begin{equation}
    \Gamma^{\alpha}_{\beta\gamma} = {1\over2}g^{\alpha\mu}\left(
    g_{\mu\beta,\gamma}+g_{\mu\gamma,\beta}-g_{\gamma\beta,\mu}\right).
\end{equation}
(Recall that ${}_{,k}$ represents the derivative $\partial/\partial
x^{k}$.)  Since the test mass is initially at coordinate rest the
$dx^{i}/d\tau$ vanish initially; so, the only connection component of
interest is $\Gamma^{i}_{tt}$.  In TT-gauge, however, this component
of the connection is identically zero (recall that $h^{TT}_{\mu\nu}$
is purely spatial); so, the equations of motion reduce to
\begin{equation}
    {d\over d\tau}{dx^{i}_{A}\over d\tau} = 0;
\end{equation}
{\em i.e.,} a free test particle at (TT-gauge) coordinate rest remains
at coordinate rest.  This applies equally well for a second component
of the detector, located at $\vec{x}_{B}$: it, too, remains at
coordinate rest.

This may seem paradoxical: if the coordinate separation of any two
components of a detector remain unchanged by the passage of
a gravitational wave, what is there to show the wave's existence? The 
paradox vanishes when we realize that coordinate separation is not 
physical separation. To determine the physical separation of the 
detector's components we must invoke the metric again. Let the 
coordinate separation between the two components of the detector at 
time $t$ be the infinitesimal
$dx^{i}_{AB}$,
\begin{equation}
    dx^{i}_{AB} = x_{B}^{i}-x_{A}^{i}.
\end{equation}
(Of course, $dt=0$ since we are talking about separation at the same 
coordinate time.)
The physical distance between these two neighboring points in spacetime is 
\begin{eqnarray}
    ds^{2} &=& g_{\mu\nu}dx^{\mu}dx^{\nu}\\
    &=& \eta_{\mu\nu}dx^{\mu}_{AB}dx^{\nu}_{AB} + 
    h^{TT}_{\mu\nu}dx_{AB}^{\mu}dx_{AB}^{\nu}\\
    &=& \eta_{jk}dx^{j}_{AB}dx^{k}_{AB} + 
    h^{TT}_{jk}dx^{j}_{AB}dx^{k}_{AB}.\label{eq:ds2TT}
\end{eqnarray}
The second term in equation \ref{eq:ds2TT} shows the effect of the
gravitational wave on the separation between the two elements of the
detector: as $h_{\mu\nu}$ oscillates, so does the distance.  If the
equilibrium separation between the components is $L$ in the direction
$\hat{n}^{j}$, to ${\cal O}(h)$ the net change $\delta L$ in the
separation is equal to
\begin{equation}
    \delta L = {1\over2}Lh^{TT}_{jk}\hat{n}^{j}\hat{n}^{k}.\label{eq:hL}
\end{equation}
The physical distance between detector components does change, in an
amount proportional to the undisturbed separation and the wave
strength as projected on the separation.  Gravitational wave detectors
are designed to be sensitive to this displacement of their components.

As mentioned above, the TT-gauge conditions amount to
eight constraints on the ten otherwise independent components of the
(symmetric) $h_{\mu\nu}$.  There are thus two components of
$h_{\mu\nu}$ that are independent of the choice of coordinate system;
correspondingly, in general relativity there are two dynamical degrees
of freedom of the gravitational field.  To see what what these amount
to, without loss of generality consider a plane wave propagating in
the $z$ direction.  Then we can write
\begin{equation}
    h:^{TT}_{\mu\nu}dx^{\mu}dx^{\nu} = 
    h_{+}(x^i,t)\left(dx^2 - dy^2\right) + 
    2h_{\times}(x^j,t)dx\, dy,
\end{equation}
where $h_{+}$ and $h_{\times}$ are the two independent dynamical 
degrees of freedom, or polarizations, of the gravitational radiation 
field. 

Solutions to the wave equation for $h_{ij}$ (eq.\ \ref{eq:wave-eq})
can be analyzed in a slow motion expansion in exactly the same way as
solutions to the Maxwell equations.\cite{jackson75a,thorne80a,finn85a}
The radiative $h_{ij}$ in this expansion divide neatly into two
classes of multipolar fields, which are (in analogy with
electromagnetism) termed {\em electric\/} and {\em magnetic\/}
multipoles.  The electric multipolar radiative fields are generated by
time-varying multipole moments of the source matter density in the
same way that the analogous electric moments of the Maxwell field are
generated by the time-varying moments of the electric charge density. 
Similarly, the magnetic radiative moments are generated by the
time-varying multipole moments of the matter momentum density, which
is the analog of the electric current density.

In electromagnetism, the first radiative moment of a charge
distribution is a time-varying charge dipole moment.  When electrical
charge is replaced by gravitational charge --- {\em i.e.,} mass --- we
see that the corresponding dipole is just the position of the system's
center of mass, which (owing to momentum conservation) is
unaccelerated.  Consequently, in general relativity there is no
gravitational dipole radiation.  The first gravitationally radiative
moment of a matter distribution arises from its ``accelerating''
quadrupole moment.  Dotting the {\em i}'s and crossing the {\em t}'s,
we find that, at leading order, the radiation field at a distant
detector is related to the matter distribution of the source according
to
\begin{eqnarray}
    h^{TT}_{ij}(t,\vec{x}) &=& {2\over r}{G\over c^{5}}
    {d^{2}\over dt^{2}}{Q}_{ij}^{TT}(t-r)\\
    Q_{ij}^{TT} &=& P_{ik}(\vec{x})Q_{kl}P_{lj}(\vec{x}) - 
    {1\over2}P_{ij}(\vec{x})Q_{lm}P_{lm}(\vec{x})\\
    Q_{ij}(t)&=& \int d^{3}x\, 
    \rho(t,\vec{x})\left(x_{i}x_{j}-{1\over3}\delta_{ij}\right),\\
    P_{jk}(\vec{x}) &=& \delta_{jk}-x_{j}x_{k}/x^{2}.
\end{eqnarray}
The expression for $h^{TT}_{ij}$ given above is the famous
``quadrupole formula'' of general relativity, which relates the
acceleration of a source's quadrupole moment to the gravitational
radiation emitted. It is, for weak gravitational fields, the exact
analog of the more famous ``dipole formula'' of electromagnetism.

\subsection{Radiated power or energy}\label{sec:insights}

Gravitational radiation carries energy away from the radiating 
system. Important insights into gravitational radiation can be gained 
by considering the energetics of radiation sources, which we do in 
this section. 

The instantaneous power carried by the radiation is, in the usual way,
proportional to the square of the time derivative of the field
integrated over a sphere surrounding the source:
\begin{equation}
L \propto {c^5\over G}4\pi r^2 \dot{h}^2.\label{eq:propL}
\end{equation}
The ``exact''\footnote{In the context of our
  approximation of everywhere weak gravitational fields.}  expression
for the power carried away in electric quadrupole radiation is
\begin{equation}
L = {1\over5}{G\over c^5}\left<{d^3 Q_{ij}\over dt^3}{d^3 Q^{ij}\over
    dt^3}\right>, \label{eq:power}
\end{equation}
where the $<>$ indicates an average several periods of the radiation.
Note that the power depends on $Q_{ij}$ and {\em not\/} $Q^{TT}_{ij}$.
  
If we focus on the radiation emitted by a weak-field, dynamical
source, we can use the multipolar expansions described above to
replace the fields by the multipole moments of the source.  For an
electric $\ell$-pole field radiated by a matter source with mass $M$,
typical dimension $R$ and internal velocity $V$,
\begin{equation}
L^{(\ell)}_{\mbox{electric}} \propto 
{c^5\over G}\left[{GM\over c^2 R}\left({V\over
    c}\right)^{\ell+1}\right]^2; 
\end{equation}
similarly, for a magnetic $\ell$-pole field
\begin{equation}
L^{(\ell)}_{\mbox{magnetic}} \propto 
{c^5\over G}\left[{GM\over c^2 R}\left({V\over
    c}\right)^{\ell+2}\right]^2
\end{equation}
The total power radiated is the sum over the power radiated in each of
the multipoles. 

Aside from numerical factors and symmetries, power radiated in the
electric $\ell$-pole channel is suppressed relative to that in the
electric quadrupole channel by a factor of $(V/C)^{2(\ell-2)}$. 
Similarly, the radiation in the magnetic $\ell$-pole channel is
suppressed from the electric quadrupole radiation by a factor of
$(V/C)^{2(\ell-1)}$.  Consequently, sources whose internal velocities
are significantly less than the speed of light radiate principally in
the electric quadrupole channel (again, unless suppressed by
symmetries).

There is still another way of looking at the power radiated by a
gravitational radiation source.  For the gravitational wave detectors
we can hope to build all the radiation of interest is of astrophysical
origin.  Excepting only a stochastic gravitational wave background,
the radiation sources are all distinct systems whose structure or
dynamics are governed by gravity.  For these systems, judicious
application of the Virial Theorem\cite{shapiro83a} allows us to relate
the internal velocities $V$ to the depth of the gravitational
potential $GM/R$,
\begin{equation}
V^2 \sim {GM\over R}. 
\end{equation}
Thus, for astrophysical sources
\begin{eqnarray}
L^{(\ell)}_{\mbox{electric}} &\propto& 
  {c^5\over G}\left(V\over c\right)^{2(\ell +3)} \simeq
  {c^5\over G}\left(GM\over c^{2}R\right)^{\ell+3}\\
L^{(\ell)}_{\mbox{magnetic}} &\propto& 
  {c^5\over G}\left(V\over c\right)^{2(\ell + 4)} \simeq
  {c^5\over G}\left(GM\over c^{2}R\right)^{\ell+ 4}.
\end{eqnarray}
Strong gravitational wave sources thus have strong internal 
gravitational fields. 

Finally, dimensional analysis of equation \ref{eq:power} for the power
radiated in the electric quadrupole leads to an important physical
insight.  Dimensionally, the system's quadrupole moment is
proportional $MR^2$.  In a closed, radiating system there is a typical
time scale $T$ for motion within the source; consequently, the total
radiated power can be written
\begin{equation}
L\propto {MR^2/T^3\over c^5/G}{MR^2\over T^3}.\label{eqn:dimAnal}
\end{equation}
The quantity $MR^2/T^2\simeq MV^2$ can be interpreted as the kinetic
energy of source matter {\em engaged in motion associated with a
time-varying quadrupolar moment.} Similarly, we identify $MR^2/T^3$ as
the instantaneous power {\em available to be radiated.} Not all this
power is radiated, however.  Equation \ref{eqn:dimAnal} shows that the
{\em fraction\/} of the available power actually radiated is equal to
the available power divided by a ``fundamental power'' defined by the
physical constants $G$ and $c$:
\begin{equation}
{c^5\over G} = 3.6\times10^{59}\,\mbox{erg/s}.
\end{equation}

The magnitude of this fundamental power gives us a feeling for the
weakness of the gravitational interaction.  For a source to radiate
even one part in $10^9$ of the power available to be tapped by the
radiation field, it must have internal motions where the kinetic
energy involved in quadrupolar motion is greater than
$3.6\times10^{50}$~erg/s. For scale, this is $10^{27}$ times greater
than the power liberated in all of the nuclear reactions occurring in
the Sun!

\subsection{Signal Power Spectrum}\label{sec:spectrum}

Observations of gravitational wave signals are always of finite
duration: either the signal is a burst of duration less than the
observation period or the signal duration is determined by the period
between when the detector is turned on and when it is turned off. A
useful characterization of this {\em observed signal\/} is its {\em
  spectrum:} the contribution to the overall mean square signal
amplitude owing to its Fourier components at a given frequency.

For definiteness, focus attention upon some particular polarization
$h(t)$ of a gravitational-wave signal that is observed over a period
$T$ beginning at $t=0$. The Fourier transform of this signal is
$\widetilde{h}(f)$:\footnote{We use the engineering convention for the
  Fourier transform.}
\begin{equation}
\widetilde{h}(f) = \int_0^T dt\,e^{-2\pi i f t} h(t).
\end{equation}
The signal spectrum is evaluated for positive frequencies and is twice
the square modulus of its Fourier transform averaged over the
observation, or
\begin{equation}
P_{h}(f) := {2\over T}|\widetilde{h}(f)|^2 
\end{equation}
for non-negative $f$. 
Since $h(t)$ is real, we can use The Parseval Theorem to obtain
\begin{equation}
\int_{0}^\infty df\,P_{h}(f) = {1\over T}\int_0^T dt\, |h(t)|^2 = 
\left<{h^{2}}\right>,
\end{equation}
where $\left<\cdot\right>$ denotes a time average. 
The signal spectral density is thus a measure of the contribution to
the mean-square signal amplitude owing to its Fourier components in a
unit bandwidth.  (For non-burst --- {\em i.e.,} stochastic or periodic ---
signals, we often take the limit as $T\rightarrow\infty$.)

As we have described it, the signal spectrum is derived from the
signal waveform $h(t)$ by ``throwing away'' the phase information.
There is clearly much less information in $P(f)$ than in the
corresponding $h(t)$: why, then, is $P(f)$ an interesting
characterization of a signal?

One reason is that real detectors are only sensitive to radiation in a
limited bandwidth --- {\em i.e.,} at certain frequencies.  The
integral of the signal power spectrum over the detector bandwidth is
the contribution to the mean-square amplitude of $h$ from power in the
detector bandwidth.

A second reason is that it is not always possible to determine the
waveform of a gravitational wave signal.  For example, the waveform of
a stochastic signal, arising from a primordial background or from the
confusion limit of a large number of weak sources, is intrinsically
unknowable.  Nevertheless, the signal spectrum is straightforward to
calculate.  In this case, the spectrum embodies everything we can know
about the gravitational wave signal.

Another example illustrates a different circumstance.  Calculations of
gravitational radiation waveforms $h_{+}(t)$ and $h_{\times}(t)$ from
the kind of stellar core collapse that triggers type II supernovae
are, even in their grossest details, extremely sensitive to the
details of the stellar model and the physics included in the
simulations.  In the face of this variety of structure, however, the
spectra all show a remarkable similarity.\cite{zwerger98a} It may be
that this variety reflects our ignorance of the relevant physics and
that with better understanding the waveforms would show much less
variation and much greater predictability; it may also be that the
details of the collapse waveform are in fact very sensitive to the
initial conditions.  Whether in practice or in principle, the waveform
is today unknown; nevertheless, the spectrum does appear to characterize
the signal quite well.

We close with a final reason that the spectrum is a useful
characterization of a gravitational wave signal.  The sensitivity of a
gravitational wave detector is limited by the detector noise, which is
an intrinsically stochastic process.  In the best detectors, the noise
is fully characterized by {\em its\/} spectrum (cf. 
\ref{sec:detNoise}).  We expect intuitively that a signal is
detectable only when its spectrum has greater magnitude than the the
detector noise spectrum over a sufficient range of frequencies.  This
qualitative notion finds quantitative expression in the {\em
signal-to-noise\/} ratio, which we discuss in \S\ref{sec:detNoise}
below.

\subsection{Conclusion}

For the purposes of detection, gravitational waves are usefully
characterized by their waveform or spectrum.  There are
important sources for which the explicit waveform is not known, either
because it is intrinsically unknowable, our grasp of the underlying
physics is not complete or the calculations involved in determining it
our beyond our capabilities.  In these cases, it may still be possible
to estimate the signal spectrum, which then serves to characterize it.

\section{Characterizing The Detector}\label{sec:charDet}

\subsection{Introduction}

Gravitational-wave detectors transform incident gravitational waves
into, {\em e.g.,} electrical signals that we can more easily
manipulate.  In \S\ref{sec:detectors}, we describe briefly and
schematically two of the detector technologies currently being pursued
to detect gravitational waves.  For all detectors we might
realistically imagine building the detector response is linear in the
incident radiation: {\em i.e.,} the time history of the detector
output is linearly related to the time history of the incident
radiation.  There are two aspects of this response that we must
consider: differential sensitivity to the radiation incident from
different directions, and differential sensitivity to incident
radiation of different frequencies.  The first of these is described
by the detector's {\em antenna pattern,} which we discuss in
\S\ref{sec:AntennaPattern}, and the second of these is described by
the detector's {\em response function,} which we discuss briefly in
\S\ref{sec:Response}.

The output of a gravitational wave detector {\em might\/} contain a
particular gravitational wave signal; however, it {\em always\/}
contains noise.  Detection, discussed in \S\ref{sec:charToDet} below,
involves distinguishing gravitational wave signals from detector
noise. To make this distinction we must have some characterization of
the signal (e.g., by waveform or by spectral density) and detector
noise.  How we characterize detector noise is the subject of
\S\ref{sec:detNoise}.

\subsection{Gravitational Wave Detectors}\label{sec:detectors}

\subsubsection{Acoustic Detectors}

The earliest and most mature detection technology is, conceptually,
nothing more than a high quality tuning fork.  Gravitational waves
excite the tuning fork; gravitational waves at or near the tuning
fork resonant frequency excite it into large amplitude oscillations.
The tuning fork is instrumented so that its acoustic vibrations become
electrical signals, which, when amplified, are the gravitational wave
signal.

Physically, the tuning fork is realized as one or more normal modes of
a large metal-alloy block: the fundamental longitudinal mode of a
right cylinder for the currently operating detectors, the five
quadrupole modes of a sphere or a truncated icosahedron
\cite{johnson93a} for the proposed next generation detectors.  The
choices made in the construction of the
ALLEGRO\cite{hamilton97a,mauceli96a} detector, built and operated at
the Louisiana State University, are typical for the current generation
of these right cylindrical ``bars'': diameter of 60~cm, length of 3~m,
and cast of Al5056 alloy for a total mass of 2296~Kg.

The mechanical oscillations of the tuning fork are converted into
electrical signals, which are then amplified, digitized, and otherwise
manipulated to determine whether gravitational waves are present or
absent.  In all of the high-sensitivity bar detectors operating today,
the transducer is not directly connected to the bar, but instruments a
second mechanical oscillator, of lower mass and smaller physical
dimension, that is itself coupled to the bar.  Gravitational waves
drive the bar, which in turn drives the second oscillator.  In the
process, the amplitude of the mechanical vibrations are amplified, and
it is this mechanically amplified motion that is converted into
electrical signals and further amplified, {\em etc.} The coupling of
the two mechanical oscillators splits the fundamental longitudinal
mode of the larger bar into two closely-spaced modes.  For the ALLEGRO
detector, the antenna's normal modes are at 896.8~Hz and 920.3~Hz.

At this writing there are five operating cryogenic acoustic
gravitational wave detectors:
\begin{itemize}
\item ALLEGRO, at the Louisiana State University in the United States
 \cite{mauceli96a},
\item AURIGA, at the University of Padua in Italy,
\item EXPLORER, operated by the Rome group and located at CERN\cite{astone97a},
\item NAUTILUS, operated by the Rome group and located at the Frascati INFN
  Laboratory\cite{astone97a}, and 
\item NIOBE, at the University of Western Australia\cite{blair96a}.
\end{itemize}
In addition to these classical ``bar'' detectors, several spherical or
truncated icosahedral detectors have been proposed or are undergoing
technical development: SFERA,
TIGA\cite{johnson93a,merkowitz95a,merkowitz96a},
GRAIL\cite{frossati97a}, and OMNI\cite{velloso97a}.

\subsubsection{Interferometric Gravitational Wave Detectors}

An alternative technology for the detection of gravitational waves
involves the use of a right-angle Michelson interferometer with freely
suspended mirrors.  Gravitational waves incident normal to the plane
of an interferometer will lead to differential changes in the distance
between the corner and end mirrors.  For frequencies much less than
the light storage time in an interferometer arm, the corresponding
motion of the fringes is proportional to the incident radiation
waveform.

There are currently two Km-scale interferometer projects under
construction: the French/Italian VIRGO Project \cite{bradaschia90a} and
the US LIGO Project \cite{abramovici92a}. VIRGO will consist of a
single interferometer with 3~Km long arms situated just outside of
Pisa, Italy. LIGO will consist of two separate facilities, one at
Hanford, Washington and one in Livingston, Louisiana. Each LIGO
facility will house an interferometer with arms of length 4~Km; in
addition, the Hanford facility will also hold an interferometer of
2~Km arm length in the same vacuum system.

In addition to these larger interferometer projects, there are three
more interferometric detectors of somewhat smaller scale under
construction: the Australian ACIGA project, the German/U.K. GEO~600
project \cite{danzmann92a} and the Japanese TAMA~300 project.  The
ACIGA Project's ultimate goal is a multi-kilometer detector, to be
located several hours outside of Perth; presently, they are beginning
the construction of an approximately 80~m prototype at the same site. 
GEO~600, located in Hanover, Germany, is a folded Michelson
interferometer with an optical arm length of 1.2~Km.  The Japanese
TAMA~300 is a 300~m Fabrey-Perot interferometer located just outside
of Tokyo; it is hoped that the success of this project will lead to 
the construction of the proposed Laser Gravitational Radiation Telescope 
(LGRT), which would be located near the Super-K neutrino detector. 

There are several ways to make an interferometer more sensitive at
frequencies less than the reciprocal of the detector's light storage
time.  One is to increase its arm length (recall equation
\ref{eq:hL}!).  The Laser Interferometer Space Antenna --- LISA --- is
an ambitious project to place in solar orbit a constellation of
satellites that will act as an interferometric gravitational wave
detector.\cite{bender96a,folkner98a} The arm length of this
interferometer would be $5\times10^6$~Km.  The LISA project has been
approved by the European Space Agency as part of its Horizon 2000$+$
Program; additionally, the US National Aeronautics and Space
Administration is actively considering joining ESA as a partner to
accelerate the development and launch of this exciting project.

\subsection{Antenna Pattern}\label{sec:AntennaPattern}

Gravitational wave detectors respond linearly to the applied field.
The interferometric and bar gravitational wave detectors now under
construction or in operation have only a single ``gravitational wave''
output channel.\footnote{Some proposed acoustic detectors are
  instrumented on several independent modes. In this case, each mode
  may be considered a separate detector and represented as a single
  gravitational wave channel.} When a plane wave is incident on such a
detector, the time history of the output channel is linearly related
to a superposition $h(t)$ of the $+$ and $\times$ polarizations of the
incident plane wave:
\begin{equation}
h = F_{{}+{}}h_{{}+{}} + F_{\times} h_{\times}.
\end{equation}
The factors $F_{{}+{}}$ and $F_{\times}$ describe the detector's
``antenna pattern'', or differential sensitivity to radiation of
different polarizations incident from different directions.  (In fact,
the antenna pattern may also be a function of radiation wavelength;
however, when the wavelength is much larger than the detector this
dependence is insignificant.)  They depend on relative
orientation of the plane-waves propagation direction and the
definition of the $+$ and $\times$ polarizations

If we fix the propagation direction and rotate the polarization of
the incident radiation, then the detector response $h(t)$ will
change. Define the polarization averaged root-mean-square (RMS) 
antenna pattern $F$,
\begin{equation}
F^{2}(\vec{k}) = \overline{F_{{}+{}}^2(\vec{k}) +
  F_{\times}^2(\vec{k})},\label{eq:F}
\end{equation}
where $\vec{k}$ is the wave-vector of the incident plane wave and the
overline denotes an average over a rotation of the incident radiation's
polarization plane.  The result depends only on the wave-vector (or,
alternatively, the wave's propagation direction and wavelength) and is
proportional to the detector's root-mean-square response
to plane-wave radiation incident from a fixed direction at fixed
wavelength.  For either the acoustic or interferometric detectors now
operating or under constructions, $F(\vec{k})$ is independent of the
magnitude of $\vec{k}$ as long as the radiation wavelength is much
larger than the detector.

A convenient pictorial representation of the detector's response
results if we plot the surface defined by $\hat{n}F(\vec{k})$ for
fixed $|\vec{k}|$, where $\hat{n}$ is the unit vector in the direction
of the source relative to the detector ({\em i.e.,}
$\hat{n}=-\vec{k}/k$).  In such a figure, the response of the detector
to a plane wave with wave-vector $\vec{k}$ (appropriately averaged
over polarization) is proportional to the distance of the surface from
the origin in the direction of the source ($\hat{n}$).  In the
remainder of this subsection we describe the antenna pattern of
interferometric and acoustic bar detectors to incident gravitational
plane waves.

\subsubsection{Bar detectors.}
In a classic bar detector, incident gravitational waves drive the
fundamental longitudinal mode of a right cylindrical bar.  The driving
force --- and thus the radiation --- is determined by observing the
motion of this mode.  For definiteness, let the longitudinal axis of
the bar be along the $\hat{z}$-direction, and consider a plane
gravitational wave propagating in the $\hat{x}$-direction:
\begin{equation}
h^{TT}_{ij}(t)dx^i dx^j = h_{{}+{}}(t)\left(dy^2 - dz^2\right)
+ 2h_{\times}(t)dy\,dz.
\end{equation}
The $+$ polarization mode changes the $z$-distance between the atoms
in the bar.  This change is resisted by inter-atomic forces in the
bar; thus, the bar's longitudinal normal mode is driven by this
polarization component of the incident wave.  The $\times$
polarization, on the other hand, does not excite the bar's mode in
this way; so, the detector is insensitive to this component of the
incident radiation.

If, on the other hand, the waves are incident along the {\em longitudinal\/}
axis, then neither the $\times$ nor the $+$ polarization components
cause any change in the longitudinal distance between the atoms in the
bar; correspondingly, the bar is insensitive to waves incident along
on the bar along its axis.

Finding the response to radiation of different polarizations incident
from directions intermediate between these extremes is a relatively
straightforward exercise in geometry.  First define the bar's
coordinate system.  Let the bar's symmetry axis define the
$\hat{z}$ direction and choose $\hat{x}$ and $\hat{y}$ such that
($\hat{x}$, $\hat{y}$, $\hat{z}$) defines a right-handed coordinate
system. 

\begin{figure}
    \epsfxsize=0.8\columnwidth
    \begin{center}
        \leavevmode\epsffile{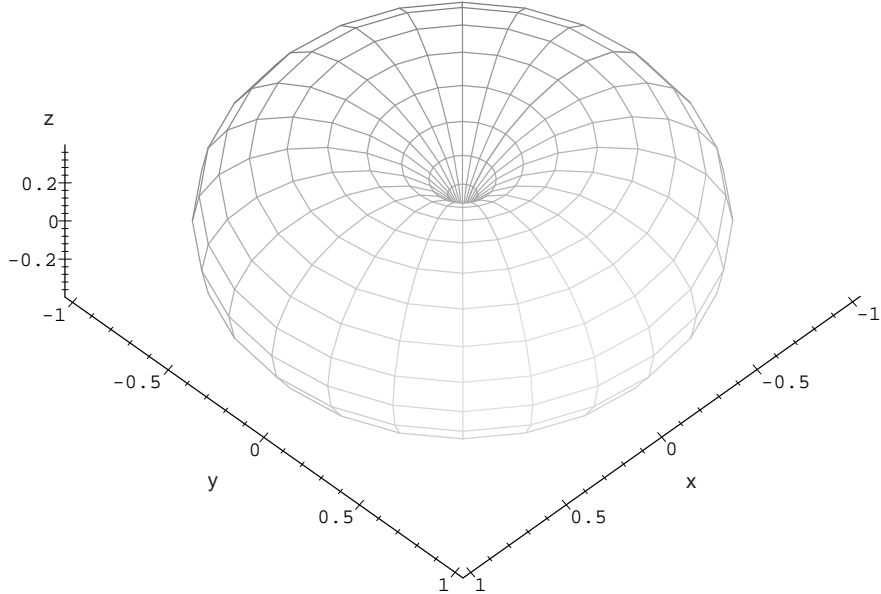}
    \end{center}
\caption{The polarization-averaged RMS sensitivity of a bar detector
  to gravitational waves incident from any direction. The detector is
  at the origin of the figure and has its symmetry axis along the
  figure's $z$ axis. The magnitude of the distance from the origin to
  the surface in a direction $\hat{n}$ is proportional to the relative
  response of the detector to radiation incident on the detector from
  that direction.}\label{fig:BarAntennaPattern}
\end{figure}

Next consider a plane gravitational wave propagating in the 
$\hat{k}$ direction and define the polarizations of the gravitational wave. 
Introduce a plane orthogonal to $\hat{k}$.  In this plane,
spanned by the Cartesian coordinates $\hat{x}'$ and $\hat{y}'$, the 
TT-gauge metric perturbation can be written
\begin{equation}
h_{ij}dx^i dx^j = h_{+}\left({dx^{\prime}}^2 - {dy^{\prime}}^2\right)
+ 2h_{\times}dx^{\prime} dy^{\prime}.
\end{equation}
The choice of $\hat{x}'$ and $\hat{y}'$ is arbitrary: different
choices correspond to either or both a rotation of one polarization
state into another or a reflection that flips the sign of
$h_{\times}$.  For definiteness, we choose $\hat{x}'$ to be orthogonal
to $\hat{z}$ and $\hat{y}'$ such that ($\hat{x}'$, $\hat{y}'$,
$\hat{k}$) is a right-handed coordinate system.  (In the degenerate
case, where $\hat{k}$ is parallel to $\hat{z}$, we choose $\hat{x}'$
parallel to $\hat{x}$ and $\hat{y}'$ parallel to $\hat{y}$.)  With
these choices, it is straightforward to show that the response of the
bar's longitudinal mode is proportional to $h$, where
\begin{eqnarray}
h &=& F_{{}+{}} h_{{}+{}} + F_{\times} h_\times\\
F_{{}+{}} &=& \sin^2\theta\\
F_{{}\times{}} &=& \sin^2\theta\\
\cos^2\theta &=& \left(\hat{k}\cdot\hat{n}\right)^2
\end{eqnarray}

Figure \ref{fig:BarAntennaPattern} shows the
polarization-averaged RMS sensitivity of a right cylindrical acoustic
detector to plane waves incident from a given direction.  To interpret
the figure, imagine a detector at the figure's origin with its
symmetry axis coincident with the figure's $z$ axis.  A plane wave,
arriving from direction $\hat{n}$, leads to a detector response
proportional to {\em the distance in the direction $\hat{n}$ from the
figure's origin to the surface.}

\subsubsection{Interferometric detectors.} 
Interferometric gravitational wave detectors respond when incident
gravitational waves cause a differential change in the length of the
interferometer arms.  Focus attention on interferometers whose arms
meet in a right angle.  To get a sense of the differential sensitivity
of such a detector to radiation of different polarizations incident
from different directions, define a right-handed interferometer
coordinate system whose origin is the intersection of the arms and
whose $x$ and $y$ coordinate directions are in the direction of the
arms.  Let a plane wave, described by the perturbation
\begin{equation}
h_{ij}dx^{i}dx^{j} = 
h_{{}+{}}\left(dx^2 - dy^2\right) + 
2h_{\times}dx\,dy,
\end{equation}
be incident on the detector from direction $\hat{z}$. There will be
no detector output proportional to $h_{\times}$, since that
component of the radiation does not lead to a differential change in
the arm lengths; on the other hand, the polarization component
proportional to $h_{{}+{}}$ does lead to a differential change in the
arm lengths and, correspondingly, to detector output.

Similarly, consider radiation incident on the detector along the
interferometer's $x$ arm:
\begin{equation}
h_{ij}dx^i dx^j = h_{+}\left(dy^2 - dz^2\right)
+ 2h_{\times}dy\,dz.
\end{equation}
Again, the $\times$ polarization mode does not lead to a differential
change in the interferometer arm lengths (at first order in $h$); so,
the detector is not sensitive to radiation with this polarization.  On
the other hand, radiation in the ${}+{}$ polarization mode, as we have
defined it, leads to changes in the length of the $y$ arm while
leaving the $x$ arm length unchanged; consequently, there is a {\em
differential\/} change in the interferometer arm length and the
detector is sensitive to radiation of this polarization incident from
this direction.

\begin{figure}
    \epsfxsize=0.8\columnwidth
    \begin{center}
        \leavevmode\epsffile{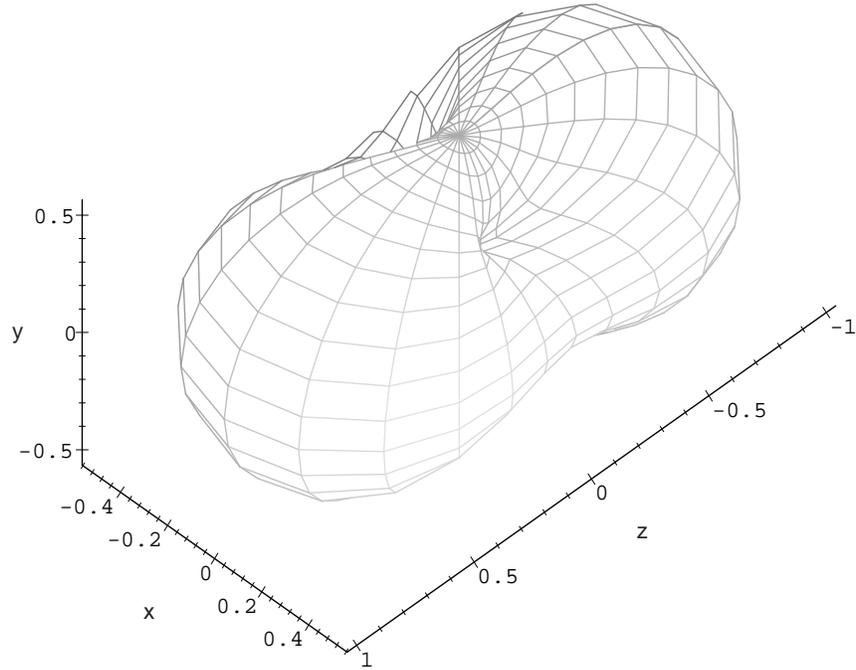}
    \end{center}
\caption{The polarization-averaged RMS sensitivity of an
  interferometric gravitational wave detector to radiation incident
  from any direction. The detector is at the origin of the figure and
  has its arms aligned with the figure's $x$ and $y$ axes. The
  magnitude of the distance from the origin to the surface in a
  direction $\hat{n}$ is proportional to the relative response of the
  detector to radiation incident on the detector from that direction,
  averaged over all polarizations.}\label{fig:IFOAntennaPattern}
\end{figure}

To determine in general the coefficients $F_{{}+{}}$ and $F_{\times}$ that
describe the response of an interferometric detector to incident plane
waves, first describe the polarization modes of radiation incident
on the detector relative to the detector coordinate system.  In the
usual ($\theta$, $\phi$) spherical coordinates associated with the
interferometer coordinate system, the incident direction of a
plane-wave propagating with wave-vector $\vec{k}$ is
\begin{eqnarray}
\cos\theta &\equiv& - \vec{k}\cdot\vec{z}/|\vec{k}|,\\
\tan\phi &\equiv& {\vec{k}\cdot\vec{y}\over\vec{k}\cdot\vec{x}}. 
\end{eqnarray}
In the plane orthogonal to the radiation propagation direction
$\hat{k}$, let the $\hat{x}'$ direction be parallel to the $xy$-plane
and the $\hat{y}'$ direction be orthogonal to $\hat{x}'$ so that
$(\hat{x}', \hat{y}', -\hat{k})$ forms a right-handed coordinate
system.  [In the degenerate case --- radiation propagating parallel to
the $\hat{z}$ direction --- we take $\hat{x}'$ parallel to $\hat{x}$
and $\hat{y}'$ such that $(\hat{x}',\hat{y}',-\hat{k})$ is
right-handed.]  In terms of this coordinate system, define the $+$ and
$\times$ polarizations of an incident gravitational wave by
\begin{equation}
h_{ij}dx^i dx^j = h_{+}\left({dx^{\prime}}^2 - {dy^{\prime}}^2\right)
+ 2h_{\times}dx^{\prime} dy^{\prime};
\end{equation}
then, the antenna pattern factors $F_{{}+{}}$ and $F_\times$ are given
by 
\begin{eqnarray}
F_{{}+{}} &\equiv& 
{1\over2}\left(1+\cos^2\theta\right)\cos2\phi\cos2\psi
- \cos\theta\sin2\phi\sin2\psi, \\
F_\times &\equiv&
{1\over2}\left(1+\cos^2\theta\right)\cos2\phi\sin2\psi
+ \cos\theta\sin2\phi\cos2\psi.
\end{eqnarray}
Figure \ref{fig:IFOAntennaPattern} shows the polarization-averaged RMS
sensitivity of a right-angle interferometric detector to plane waves
incident from a given direction.  The detector is at the origin of the
figure, with its arms along the figure's $\hat{x}$ and $\hat{y}$ axes. 
The detector's sensitivity to radiation incident on the detector from
direction $\hat{n}$ is proportional to the distance of the surface
from the figure's origin in the direction $\hat{n}$.

\subsection{Response Function}\label{sec:Response}

The output of a gravitational wave detector is a voltage, $v(t)$, that
is linearly related to the incident radiation.  Consider a
gravitational plane wave, with polarizations $h_{{}+{}}$ and
$h_{\times}$, incident on a detector with antenna pattern described by
$F_{+}$ and $F_\times$.  The detector response is given by
\begin{eqnarray}
v(t) &=& \int_{-\infty}^t d\tau\, h(\tau) K(t-\tau), \qquad\mbox{where}\\
h(t) &=& F_{+}h_{+}(t) + F_{\times}h_{\times}(t)
\end{eqnarray}
and $K$ is the kernel of the linear transformation. 

It is instructive to express this convolution in the frequency domain:
\begin{eqnarray}
v(t) = \int_{-\infty}^\infty df\,\widetilde{h}(f)
\widetilde{K}^{*}(f)\exp\left(2\pi i f t\right),
\label{eq:lt}
\end{eqnarray}
where
\begin{equation}
\widetilde{g}(f) \equiv \int_{-\infty}^\infty dt\, g(t)\exp(-2\pi i f t)
\end{equation}
and we have assumed that $K(\tau)$ vanishes for negative
$\tau$.\footnote{Corresponding to a causal impulse response!} From
equation \ref{eq:lt} we see that the response of the system --- the
output voltage --- depends on the frequency of the incident radiation:
depending on the character of the detector, the response may be
relatively large
for some frequencies and relatively small for others.

As an example, consider two equal masses $M$ connected by a spring
(spring constant $\omega_0^2 M$, quality factor $Q$). Denote
the equilibrium separation by $L$. A passing gravitational wave of
appropriate polarization disturbs the equilibrium separation of the
system.  The net result is that the passage of a gravitational wave
acts as a driving force on the system's normal mode:
\begin{equation}
\ddot{x} + {\omega_0\over Q}\dot{x} + \omega_0^2 x = {1\over2}L\ddot{h},
\end{equation}
where $x$ is the difference between the actual and equilibrium
separation. 

Suppose that we instrument this system with strain gauges to
produce an output voltage $v(t)$ proportional to $x(t)$, the deviation
from equilibrium separation. How is $v(t)$ ({\em i.e.,} $x(t)$)
related to $h(t)$? In the 
frequency domain we see that
\begin{equation}
\widetilde{v}(f) \propto {f^2\over f_0^2-f^2 + iff_0/Q}\widetilde{h}(f),
\end{equation}
where
\begin{equation}
f_0 = \omega_0/2\pi.
\end{equation}
The output voltage for excitations near the resonance can vary
dramatically as a function of frequency.

The response function we have just described is equivalent
conceptually to that of a modern acoustic detector: the radiation
manifests itself as a driving force on the system's normal modes and
the response is a strong function of the frequency in the neighborhood
of the resonances.

The response function of an interferometric detector is quite
different.  For an interferometer, at frequencies much below the
round-trip travel time (but greater than the pendulum frequency of the
suspended mirrors and beam-splitter) the detector response is
independent of frequency; only when the frequency becomes comparable
to or larger than the round-trip light travel time in an
interferometer arm does the response vary with
frequency.\cite{saulson97a}\footnote{It is commonly said that an
interferometer responds to a passing gravitational wave
proportionately with the differential change in the IFO arm length. 
This is not quite right.  The response of an interferometric detector
to a passing gravitational wave is proportional to the differential
change in the round-trip light travel time in the arms.  The
round-trip light travel time involves the integrated change in the arm
length over the past, as opposed to the instantaneous separation at
the time of reflections.  For frequencies small compared to the
inverse round-trip light travel time the difference is negligible.
  
It is also the case for interferometers that the frequency dependence
of the response function varies with the incidence direction of the
radiation though --- again --- this is only significant at frequencies
comparable to or larger than the inverse round-trip light travel time
in an arm.}

The amplitude of the response determines those frequencies where an
incident gravitational wave of unit amplitude gives relatively large
amplitude output and where it gives relatively small amplitude output.
It is not, however, the case that relativity large amplitude output
corresponds to relatively large {\em sensitivity,\/} if by sensitivity
we mean greater ability to detect.  To address the question of
sensitivity, we must turn to yet a different aspect of a detector's
function: its noise.

\subsection{Noise}\label{sec:detNoise}

The output channel of a gravitational wave detector is always alive
with random fluctuations --- {\em noise\/} --- even in the absence of
a gravitational wave signal.  In a perfect world noise would arise
exclusively from fundamental physical processes: {\em e.g.,}
fluctuations owing to the finite temperature of the detector, counting
statistics of individual photons on a photo-detector, {\em etc.} In the
less than perfect world in which we live there will be other
contributions to the detector noise, beyond these fundamental
processes, that arise from the imperfect construction of the detector
({\em e.g.,} bad electrical contacts), imperfections in the materials
used to construct the detectors ({\em e.g.,} mechanical creep and strain
release), and from the detector's interaction with the
(non-gravitational wave) environment ({\em e.g.,} seismic vibrations,
electromagnetic interactions, {\em etc.}).

Detection of gravitational waves requires that we be able to 
distinguish, in the detector output, between signal and noise. This 
requires that we have characterized the noise (and not only the signal). 
Since noise is intrinsically
random in character, that characterization is in terms of its
statistical properties. Some of these statistical properties we can
predict, model or anticipate {\em a priori,} based on the detector
design; nevertheless, it is important to realize that an experimental
apparatus is a real thing made in the real world and will never behave
ideally. While a large part of the experimental craft involves
building instruments that operate as close as possible to their
theoretical limits or prior expectations, the final characterization
of a detector will always be determined or verified empirically. In
this section we describe something of how noise in gravitational wave
detectors is characterized.

\subsubsection{Correlations}

Just as a probability distribution is fully characterized by its
moments, so the random output of a gravitational wave detector can be
fully characterized by its {\em correlations.} The $N$-point
correlation function describes the mean value of the product of the
detector output sampled at $N$ different times. Mean, in this case,
refers to an {\em ensemble\/} average, where the ensemble is an
infinite number of identically constructed detectors. Denoting by
$n(t)$ the noisy output of a gravitational wave detector in the
absence of any signal, the $N$-point correlation function of the noise
distribution is given by
\begin{equation}
C_N(\tau_0,\ldots,\tau_{N-1})
= \overline{n(\tau_0)\ldots n(\tau_{N-1})},
\end{equation}
where the over-bar signifies an ensemble average, which is also
referred to as an average {\em across the process.}

As a practical matter ensemble averages are impossible to realize
experimentally: one rarely has the opportunity of working with even
two similar detectors, let alone an infinite number of identical
ones. Thus, while a handy theoretical construct, the general set of 
correlation functions is not of great practical use in characterizing
the behavior of a real detector. 

\subsubsection{Stationarity}

If, however, the behavior of the detector noise does not depend
significantly on time --- {\em i.e.,} the noise is {\em stationary\/}
--- then the utility of the correlation function as a practical tool
for characterizing detector noise increases dramatically.  When the
noise character is, figuratively, the same today as it was yesterday
and as it will be tomorrow, then the detector yesterday (or an hour,
or a minute, or a second ago) can be regarded as an identical copy of
the detector we are looking at now, and both are identical copies of
the detector tomorrow.  Consequently, in the spirit of the ergodic
theorem, we can replace the average across the process --- the
ensemble average --- with an average {\em along\/} the process --- a
time average.  The $N$-point correlation function is then a function
of the difference in time between the $N$ samples:
\begin{equation}
C_N(\tau_1, \ldots, \tau_{N-1}) = 
\lim_{T\rightarrow\infty}{1\over T}\int_{t_0-T}^{t_0}
n(t)n(t - \tau_1)\cdots n(t-\tau_{N-1})\,dt.\label{eq:ergodic}
\end{equation}

Of course, perfect stationarity is an impossible requirement.  As a
practical matter, what we require is that the noise process be
stationary over a suitably long period.  Let's try to make that
concept more quantitative.  To simplify the discussion, assume
(without loss of generality) that the noise process has zero mean. 
Consider first the two-time correlation function of a stationary
process:
\begin{equation}
C_2(\tau) = \lim_{T\rightarrow\infty}{1\over T}\int_{t_0-T}^{t_0}
n(t)n(t - \tau)\,dt.\label{eq:average}
\end{equation}
For sufficiently large $\tau$ we expect intuitively that $C_2(\tau)$
should vanish: the output now should be effectively uncorrelated with
the output in either the distant past or the distant future.  This
will also be the case for the higher-order moments as well: for
sufficiently large $\tau_k$ (any $k$), the correlation function $C_N$
should vanish.  Thus, we don't need to require perfect stationarity;
rather, we require only that the statistical character be
approximately stationary, varying significantly only over times long
compared to the longest correlation time.  In that case, we can
approximate the correlations $C_N$ by averaging, as in equation
\ref{eq:average}, over {\em finite\/} periods.

\subsubsection{Gaussian Noise}\label{sec:GaussianNoise}

Noise from fundamental processes tends to be either Gaussian ({\em
  i.e.,} originating from contact with a heat bath or some dissipative
process) or Poissonian ({\em e.g.,} originating in the counting
statistics of identical and independently distributed --- i.i.d. ---
events that occur at a fixed, average rate).  
For the gravitational wave detectors
under construction, the intrinsically Poissonian processes ({\em
  e.g.,} photon counting statistics) have rates so high that they can
be treated as Gaussian and we do so here and below.

One way to think about the detector noise is as a superposition of a
Gaussian and approximately stationary component, a (hopefully lower
amplitude) non-Gaussian, but still stationary component, superposed
finally with a non-stationary component. General statements cannot be
made about the non-Gaussian or non-stationary components: they differ
from instrument to instrument and environment to environment and can
only be characterized empirically. The characterization of the
Gaussian-stationary component, however, is remarkably simple and has a
useful physical interpretation, which we review in this and the next
subsection. (For more information and detail, see Finn\cite{finn92a}.)

Up to now we have considered the output of a detector as an analog
process: {\em i.e.,} one that is continuous in time. In fact, the
output we observe will have been sampled discretely at some sampling
rate $f_s$, chosen to be something more than twice as great as the
maximum frequency of interest for the detector output. So, instead of
writing the noise at the detector output as $n(t)$ we write
\begin{equation}
n[k] \equiv n(t_k), 
\end{equation}
where
\begin{equation}
t_k = t_0 + k \Delta t
\end{equation}
for constant $\Delta t$. 

When the noise in the detector is Gaussian and stationary, any single
sample $n[j]$ of the detector output is drawn from a normal
distribution with a mean and variance that are independent of when the
sample was taken.  Without loss of generality we can assume that the
mean $\overline{{n}}$ vanishes, in which case
\begin{equation}
P(n[j]) = {
\exp\left[-n[j]^2/2\sigma^2\right]
\over\sqrt{2\pi\sigma^2}
} . \label{eq:p(n[j])}
\end{equation}
We understand the variance $\sigma^2$ of the distribution to be the
ensemble average of the square of the detector noise:
\begin{equation}
\sigma^2 = \overline{n[j]^2}.
\end{equation}

Equation \ref{eq:p(n[j])} holds true for each sample $n[j]$;
consequently, the {\em joint probability\/} that the length $N_T$ {\em
  sequence\/} of samples $n[j]$, $j$ running from 1 to $N_T$, is a
sample of detector noise is given by the {\em multivariate\/} Gaussian
distribution
\begin{equation}
P(n[1],n[2],\ldots,n[N_T]) =
{\exp\left[-{1\over2}\sum_{j,k=0}^{N_T-1}
n[j]\left|\left|{\bf C}^{-1}\right|\right|_{jk}n[k]\right]\over
\sqrt{\left(2\pi\right)^{N_T}\det||{\bf C}||}}
\label{eq:p(n)}
\end{equation}
In place of the variance $\sigma^2$ that appears in the exponent of
equation \ref{eq:p(n[j])} is the {\em covariance\/} matrix
${\bf C}$. (The matrix ${\bf C}^{-1}$ that appears in equation
\ref{eq:p(n)} is, by construction, positive definite; consequently,
it is non-singular and invertible.)  Similarly, in place of the factor
$\sigma^2$ that appears in the denominator of equation
\ref{eq:p(n[j])} is the determinant $\det||{\bf C}||$.

The mean over the product $n[j]n[k]$ is the value of the correlation
function $C_{2}(t_{j}-t_{k})$; it is also just the value of the $jk$
element of the covariance matrix $\left|\left|C\right|\right|$:
\begin{equation}
    C_{2}[j-k] = \overline{n[j]n[k]} = \left|\left|{\bf C}\right|\right|_{jk}.
\end{equation}
Since the detector noise is also assumed to be
stationary, $\left|\left|{\bf C}\right|\right|_{jk}$ can depend only
on the difference $j-k$; correspondingly, ${\bf C}$ is constant on
its diagonals: {\em i.e.,} it is a {\em Toeplitz\/} matrix.
Consequently, it is fully characterized by the sequence $c[k]$ of
length $2N_T-1$ whose elements are the first row and column of
${\bf C}$:
\begin{equation}
c[j-k] =  \left|\left|{\bf C}\right|\right|_{jk} = C_2(t_j-t_k). 
\end{equation}

The sequence $c$ and the process mean 
(which we have assumed to vanish) fully characterize the random process.
The sequence $c$, however, is just the two-time correlation function
of the detector output! Thus, the two-time correlation function $C_2$
fully characterizes a Gaussian stationary process: all the higher
order correlation functions $C_N$ either vanish (for odd $N$) or are
expressible as sums of products of $C_2$. Once we have determined
$C_2$, then, we have completely determined the character of the
Gaussian noise process. 

\subsubsection{Likelihood function}\label{sec:likelihood}

In the last section we evaluated
\begin{equation}
    P(v|0) \equiv \left(\begin{array}{l}
    \mbox{probability of observing}\\
    \mbox{output sequence $v$ assuming}\\
    \mbox{no signal is present}
    \end{array}
    \right)
\end{equation}
for Gaussian-stationary detector noise. 
Since the detector is linear, the probability 
\begin{equation}
    P(v|h) \equiv \left(\begin{array}{l}
    \mbox{probability of observing}\\
    \mbox{output sequence $v$ assuming}\\
    \mbox{signal $h$ is present}
    \end{array}
    \right)
\end{equation}
is just
\begin{equation}
    P(v|h) = P(v-v_{h}|0),
\end{equation}
where $v_{h}$ is the detector response to the gravitational wave 
signal $h$. 
The ratio of these two probabilities, 
\begin{equation}
    \Lambda(v|h) \equiv {P(v|h)\over P(v|0)},
\end{equation}
termed the {\em likelihood function,} is the {\em odds\/} that the
data $v$ is a combination of signal $v_h$ and noise, as 
opposed to a noise alone.  For a given
observation $v$ the likelihood can be viewed as a function of
hypothesized signal $h$, in which case it has a convenient
interpretation in terms of {\em plausibility:} in particular,
$\Lambda(v|h)$ can be interpreted as the {\em plausibility\/} that the
signal $h$ is present given the particular observation $v$.  (The
likelihood is not, however, a probability.)  This meaning of the
likelihood is independent of the statistical character of the noise. 
The difficulty, if the noise is not Gaussian-stationary, is in
evaluating $\Lambda$.

\subsubsection{The two-time correlation function}

The correlation function $C_2(\tau)$ describes the statistical
relationship between pairs of samples drawn from the random process
$n(t)$ at times separated by an interval $\tau$. Given two samples
separated in time by $\tau$, a non-zero correlation $C_2(\tau)$
corresponds to an increased ability to predict the value of one member
of the pair given the other. 

The correlation function $C_2(\tau)$ is bounded by $\pm{}C_2(0)$,
suggesting that we define the {\em correlation coefficient}
\begin{equation}
R_2(\tau) \equiv C_2(\tau)/C_2(0),
\end{equation}
which is bounded by $\pm1$.  If the correlation coefficient is zero
for some $\tau$, then samples taken an interval $\tau$ apart are
entirely uncorrelated: knowledge of one does not lead to any increased
ability to predict the other.  A positive correlation coefficient
tells us that the two samples are more likely close to each other in
magnitude and sign than not, while a negative correlation coefficient
tells us that the two samples are likely close to each other in
magnitude but of opposite sign.  The larger the coefficient magnitude
the greater the tendency.  When the correlation coefficient is unity
then the correlation is perfect: {\em i.e.,} when it is $+1$ the two
samples are always equal, and when it is $-1$ the two samples are
always of equal magnitude but opposite sign.

\subsubsection{Noise Power Spectral Density}

Consider for a moment a simple harmonic oscillator --- {\em e.g.,} a
pendulum --- coupled weakly to a heat bath. The heat bath excites the
oscillator so that its mean energy is $k_B T$. Since the coupling to
the heat bath is weak, the phase of the oscillator progresses
nearly uniformly in time with rate $\omega_0$ corresponding to the
oscillator's natural angular frequency. Over long periods, however, the
continual, random excitations of the oscillator cause the phase to
drift in a random manner from constant rate.

Now suppose that we sample the position coordinate of the oscillator 
at intervals separated by exactly one period $2\pi/\omega_0$.  Since 
the coupling to the heat bath is weak the samples are very nearly 
identical: in fact, were it not for the contact with the heat bath, 
they would be exactly identical.  Thus, we expect that the correlation 
coefficient corresponding to an interval equal to an oscillator period 
should be nearly unity.  Continuing to focus on samples taken at 
intervals equal to exact multiples of the period, we expect that the 
correlation coefficients should remain large for small multiples, but 
should decrease as the interval increases since contact with the heat 
bath will lead, as time increases, to greater drift in the phase.

On the other hand, suppose that we sample the position coordinate of
the oscillator at intervals separated by exactly odd integer multiples
of a {\em half}-period $\pi/\omega_0$. Now we expect the correlation
coefficient to be nearly equal to $-1$ for small intervals, decreasing
in magnitude to $0$ as the interval increases. 

\begin{figure}
\epsfxsize=0.8\columnwidth
\begin{center}
\leavevmode\epsffile{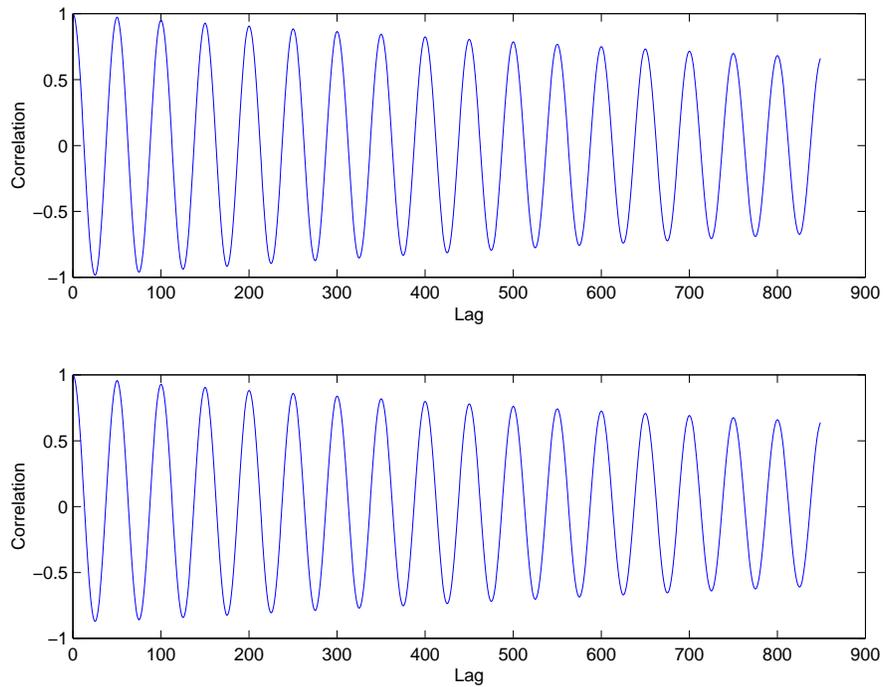}
\end{center}
  \caption{The correlation functions for a harmonic oscillator in
  contact with a heat bath.  Contact with a heat bath leads to
  damping; the nature and degree of the contact determine the
  character of the system's noise.  In this figure we show the
  correlation function, over several periods, for two different kinds
  of heat bath contact with the same on-resonance damping.  The upper
  panel corresponds to a viscous damped harmonic oscillator; the lower
  panel corresponds to a structurally damped oscillator.  The
  difference between the two correlation functions is apparently very
  subtle.}
  \label{fig:C2} 
\end{figure}

Contact with a heat bath can take place in many ways, leading to
subtly different correlation functions. Figure \ref{fig:C2} shows the
correlation function corresponding to two different kinds of heat bath
contact: that which leads to velocity damping and that which leads to
structural damping\cite{saulson90a}. Note how, pictured in this way,
there is apparently little difference between these two damping
measures.

\begin{figure}
\epsfxsize=0.8\columnwidth
\begin{center}
\leavevmode\epsffile{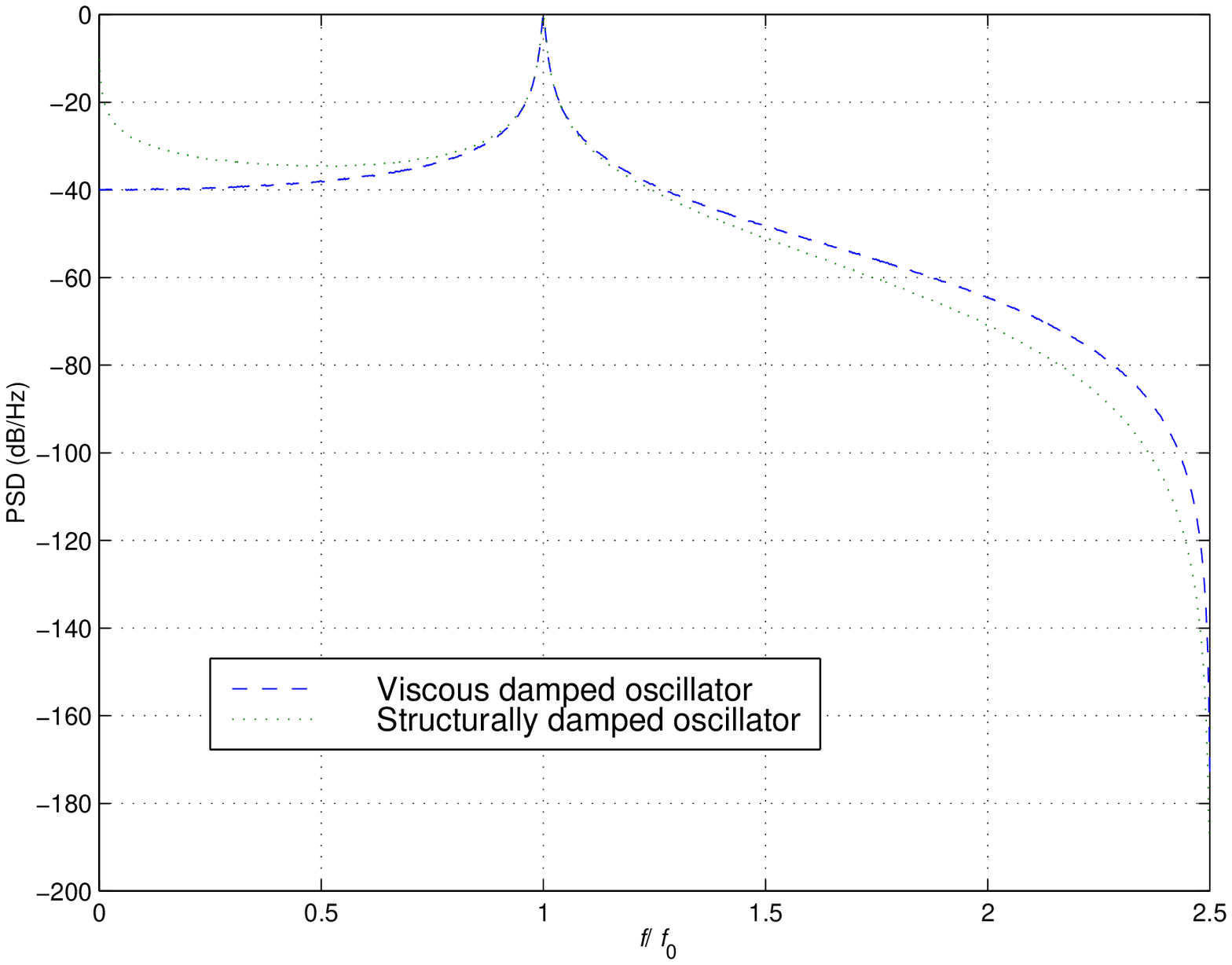}
\end{center}
\caption{The power spectral density of the two processes whose
  correlation functions are shown in figure
  \protect{\ref{fig:C2}}. Note that, while the correlation functions
  appear very similar as functions of time, strong differences show up
  in the power spectral densities as functions of
  frequency.}\label{fig:PSD} 
\end{figure}

Since the correlation function is so oscillatory we are immediately
led to consider its Fourier transform.  In this case, since
$C_2(\tau)$ is an even function of the lag $\tau$, we consider the
cosine transform, which we term the {\em one-sided power spectral
density:}
\begin{equation}
S_v(f) = 4\int_{0}^\infty d\tau\, C_2(\tau)\cos(2\pi f\tau).
\end{equation}
({\em One-sided\/} refers to the fact that, in choosing a cosine
transform, we have effectively folded the power in negative
frequencies into the power at positive frequencies; so, the $S_v(f)$
includes the power at frequencies whose magnitude is $|f|$.)  Figure
\ref{fig:PSD} shows the power spectral densities corresponding to the
correlation functions of figure \ref{fig:C2}. The strongly oscillatory
nature of these functions shows up as a large peak at the oscillator
resonant frequency (normalized to unity). In addition, however, the
PSD shows clearly the very different off-resonance character of the
noise. Noise from a structurally damped system rises in amplitude as
the frequency falls below resonance, unlike the noise contribution
from a viscously damped system; similarly, noise from a structurally
damped system falls more steeply with frequency above resonance than
does the noise from a viscously damped system.  To see the same in the
correlation function would require close inspection of the trends of
the correlation function envelopes over very long lags.

Thus, even though it is completely equivalent to the correlation
function, the power spectral density is often a more useful
characterization of the noise character.  In the case of Gaussian
noise its equivalence to the correlation function guarantees it is
also a full characterization of the detector noise.  When the noise is
not Gaussian, there are analogous spectra associated with the higher
order correlations: for example, the bispectrum is the 2-dimensional
Fourier transform of $C_3(\tau_1,\tau_2)$,
\begin{equation}
    \mbox{Bi}(f_{1},f_{2}) = 
    \int_{-\infty}^{\infty}d\tau_{1}
    e^{-2\pi i f_{1}t_{1}}
    \int_{-\infty}^{\infty}d\tau_{2}\,
    e^{-2\pi i f_{2}t_{2}}\,
    C_{3}(\tau_{1},\tau_{2}),
\end{equation}
and so on. These higher order spectra and their magnitudes play the 
same role for the higher-order correlation functions as the power 
spectral density plays for the auto-correlation function. 

\subsection{Signal-to-noise ratio}

When is a gravitational wave ``detectable''?  We haven't yet explored 
the meaning of ``detection'' qualitatively, let alone quantitatively; 
nevertheless, we have an intuitive feeling that a signal ought to be 
detectable if the detector's response to the signal is greater than 
the intrinsic noise amplitude.  Let's develop that idea a bit.

Suppose that we have a detector with noise power spectral density
$S_{v}(f)$ and particular output $v(t)$, which consists of a signal
$v_{h}(t)$ superposed with detector noise $v_{n}(t)$. The variance 
of $v(t)$, over an interval $[0,T]$, is
\begin{eqnarray}
    \sigma^{2}_{v} &=& {1\over T}\int_{0}^{T} dt\, v(t)^{2}\\
    &=& {2\over T}\int_{0}^{\infty} df\, \left|\widetilde{v}(f)\right|^{2}.
\end{eqnarray}
The noise is a random process; so, then, is $\sigma^{2}_{v}$. Focus 
on the ensemble average of $\sigma^{2}_v$ and look in the 
frequency domain:
\begin{eqnarray}
    {d\overline{\sigma^{2}_{v}}\over df} &=& 
    {2\over T} \overline{\left|\widetilde{v}(f)\right|^{2}}\\
    &=& {2\over T}\left(
    \overline{\left|\widetilde{v}_{n}(f)\right|^2} + 
    \left|\widetilde{v}_{h}(f)\right|^2
    \right)
\end{eqnarray}
where the final equality follows when we recognize that the noise is 
independent of the signal. The contribution to the mean signal 
variance thus consists of separate contributions from the signal and 
from the noise. 

The ratio
\begin{equation}
    {\left|\widetilde{v}_{h}(f)\right|^2\over
    \overline{\left|\widetilde{v}_{n}(f)\right|^2} }
\end{equation}
evidently tells us which --- signal or noise --- is expected (note
ensemble average!)  to contribute more to the amplitude of the
detector output in a unit bandwidth about frequency $f$.  We can
compute a similar, dimensionless quantity over the full bandwidth:
\begin{equation}
    \int_{-\infty}^{\infty} df\,
    {\left|\widetilde{v}_{h}(f)\right|\over
    \overline{\left|\widetilde{v}_{n}(f)\right|}/T}
    = 4\int_{0}^{\infty}df\,
    {\left|\widetilde{v}_{h}(f)\right|\over S_{v}(f)}
\end{equation}
tells us which of the signal $v_{h}$ or the noise $v_{n}$
is expected to contribute more to the variance of the output $v$. 

Given a particular sample of detector output $v$ we don't know, {\em a
priori,} what part is $v_{n}$ and what part (if any) is $v_{h}$. 
Consider a quantity that we can calculate directly from the detector
output $v$:
\begin{equation}
    \rho^{2} \equiv 
    4\int_{0}^{\infty}df\,{\left|\widetilde{v}(f)\right|^{2}\over 
    S_{v}(f)}.\label{eq:snr}
\end{equation}
The integrand is evidently the ratio of the actual contribution to the
signal variance in a unit band about frequency $f$ to the contribution
that would be expected, in the same band, from noise alone.  Not
surprisingly, the ensemble mean $\overline{\rho^{2}}$ is
\begin{equation}
    \overline{\rho^{2}} = 1 + 4\int_{0}^{\infty}df\,
    {\left|\widetilde{v}_{h}(f)\right|\over S_{v}(f)}
\end{equation}
We refer to $\rho^{2}$ as the {\em signal-to-noise ratio,} or
SNR.\footnote{Note that this definition of $\rho^{2}$ is different, by
the additive factor of unity, than used elsewhere in the gravitational
wave literature.}

Our construction of $\rho^{2}$ has been physically motivated. It turns
out, however, that exactly this same quantity arises from a
consideration of the probability $P(v|0)$, which we explored in
\S\ref{sec:GaussianNoise}.  In that section we found, for
Gaussian-stationary noise,
\begin{equation}
    P(v|0) = 
    {\exp\left[-{1\over2}
    \sum_{j,k=0}^{N-1}v[j]\left|\left|{\bf C}^{-1}\right|\right|_{jk}v[k]
    \right]\over
    \sqrt{2\pi\det\left|\left|{\bf C}\right|\right|}}
\end{equation}
With just a little algebra, however, the argument of the exponential
can be rewritten as\cite{finn99a}
\begin{eqnarray}
\sum_{j,k=0}^{N-1}v[j]\left|\left|{\bf C}^{-1}\right|\right|_{jk}v_{h}[k]
&=& 
\Re\left[{1\over2N-1}
\sum_{j=-(N-1)}^{N-1}{\left|\widetilde{V}[j]\right|^{2}\over\widetilde{c}[j]}
\right]
\label{eq:argExp}
\end{eqnarray}
where the periodic sequence $\widetilde{g}[j]$ is related to the 
discrete Fourier transform of the sequence $g[k]$:
\begin{equation}
    \hat{g}[j] = \sum_{k=-(N-1)}^{N-1}
    e^{-2\pi i kj/(2N-1)}{g}[k]\label{eq:dft}
\end{equation}
and $V[k]$ is just $v[k]$ zero-padded for negative $k$:
\begin{equation}
    V[k] = \left\{\begin{array}{ll}
    v[k]&\mbox{for $k\geq0$}\\
    0&\mbox{for $k<0$}.
    \end{array}
    \right.
\end{equation}
These summations can be regarded as approximations to integrals, in
which case
\begin{equation}
\Re\left[{1\over2N-1}
\sum_{j=-(N-1)}^{N-1}{\left|\widetilde{V}[j]\right|^{2}\over\widetilde{c}[j]}
\right]\propto\int_0^\infty df\,{\left|\widetilde{V}\right|^2\over S_h(f)}.
\end{equation}
Hence, the SNR associated with the observed detector output $v$ is
closely related to the probability that $v$ is a sample of {\em
  just\/} detector noise, with no gravitational-wave signal present.
The larger $\rho^{2}$, the smaller this probability. Should we observe
detector output with large $\rho^{2}$, then, we are not too far wrong
to be suspicious that we have seen evidence for gravitational waves.

To make this last judgment --- which involves making quantitative 
the notion of ``large'' $\rho^{2}$ --- we need to know the probability
distributions of the SNR in both the presence and absence of a signal:
after all, since noise is a random process there is some non-zero
probability that, in any given observation, $\rho^{2}$ will take on
any particular value, large or small. We return to consider this 
point in \S\ref{sec:charToDet}

\subsubsection{Matched Filtering}
\begin{figure}
    \epsfxsize=0.8\columnwidth
    \begin{center}
        \leavevmode\epsffile{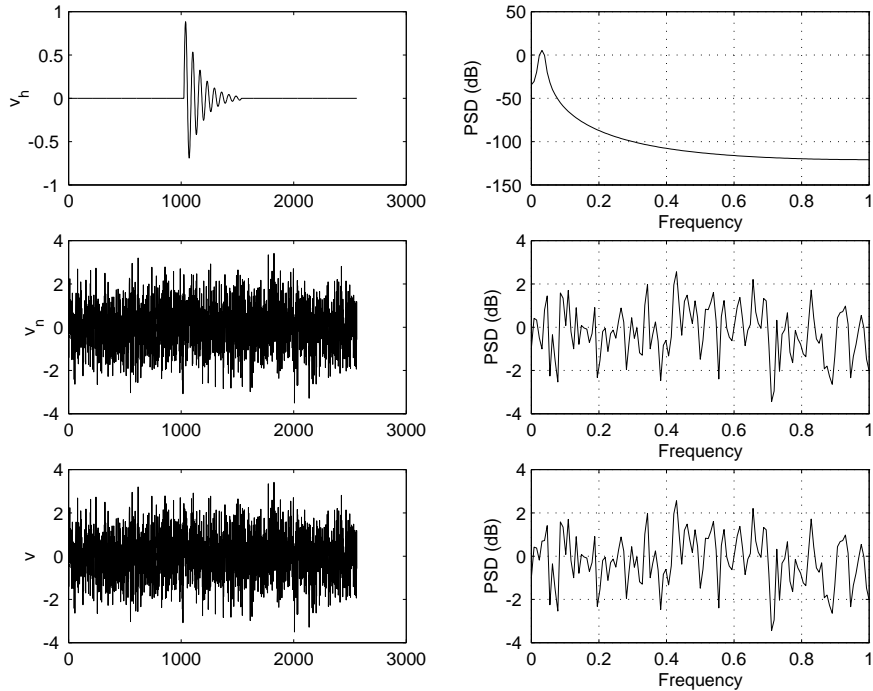}
    \end{center}
    \caption{An imagined gravitational wave signal (upper-left
    panel), detector noise (middle-left panel) and total detector 
    output (signal$+$noise, lower-left panel). Note how the signal 
    is not evident to the eye in the detector output. The right-hand 
    panels show the power spectra of the corresponding left-hand panels; 
    again, the signal power is not evident relative to the noise 
    power.}\label{fig:mfin}
\end{figure}

Calculating $\rho^{2}$ defined by equation \ref{eq:snr} does not
require or make use of any information about the gravitational
radiation source.  Suppose that we know, {\em a priori,} the radiation
waveform has the shape $V_h(t)$, and that the question is whether the
corresponding signal $\alpha V_{h}(t-t_0)$, for some unknown constants
$\alpha$ and $t_0$, is present in the detector observed output $v(t)$.
Can we make use of this information --- the signal shape $V_{h}(t)$
--- to boost our ability to observe the signal?

\begin{figure}
    \epsfxsize=0.8\columnwidth
    \begin{center}
        \leavevmode\epsffile{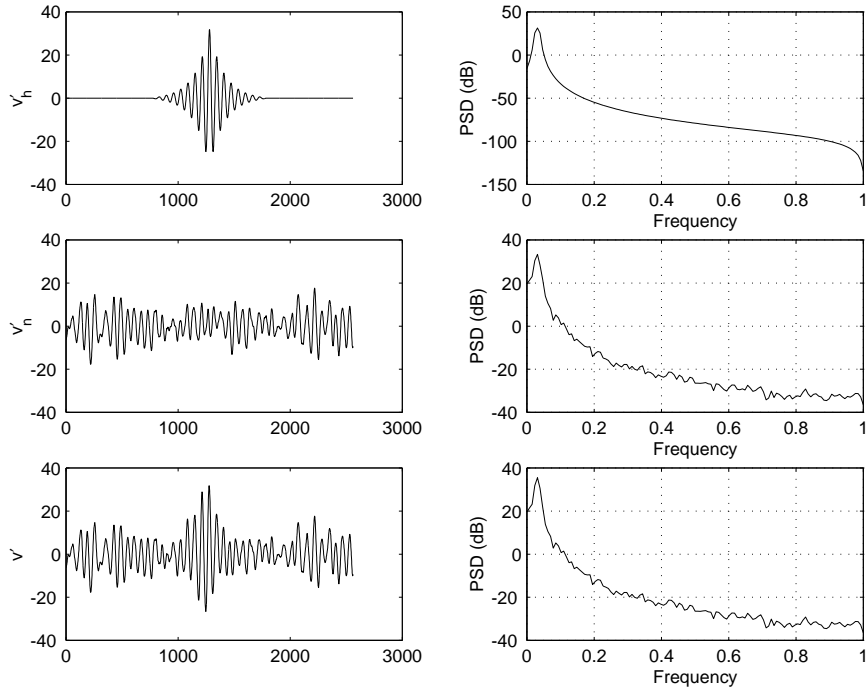}
    \end{center}
    \caption{The output of the filter described in equation
    \protect{\ref{eq:v'}} when just the signal $v_{h}$ is filtered
    (upper panel) and when the detector output, consisting of signal
    and noise, is filtered (lower panel).  In contrast to the
    lower-right panel of figure \protect{\ref{fig:mfin}}, the
    ``signal'' ({\em i.e.,} the upper panel) is quite evident even in
    the presence of noise.}\label{fig:mfout}
\end{figure}

The answer is yes.  To illustrate, figure \ref{fig:mfin} shows an
imagined $v_{h}$, $v_{n}$ and $v$ equal to $v_{h} + v_{n}$ in the
left-hand panels, and the corresponding power spectra in the
right-hand panels.  For this illustration we have assumed that the
noise is white across the detector bandwidth.  The signal is not
apparent to the eye in either $v$ or its power spectrum $P_{v}(f)$.
Figure \ref{fig:mfout} shows, in the top panel, the filter output 
when just $v_{h}$ is passed through the filter $K$ with impulse-response 
$V_{h}$ set equal to $v_{h}$:
\begin{eqnarray}
    v'(t) &=& \int_{-\infty}^{t}d\tau\, v(\tau)K(t-\tau)\\
    &=& v'_{n}(t) + v'_{h}(t)\label{eq:v'}
\end{eqnarray}
where
\begin{eqnarray}
    K(\tau) &=& v_{h}(t)\\
    v'_{n}(t) &=& \int_{-\infty}^{t}d\tau\,v_{n}(\tau)V_{h}(t+\tau)\\
    v'_{h}(t) &=& \int_{-\infty}^{t}d\tau\,v_{h}(\tau)v'_{h}(t+\tau).
\end{eqnarray}
Without loss of generality we assume $v_{h}$ is non-zero only for
positive $t$.  The filtered detector output $v'(t)$ consists of a
signal contribution $v'_{h}(t)$ and a noise contribution $v'_{n}(t)$.
These are shown in the top and middle panels of figure
\ref{fig:mfout}, respectively.  The bottom panel of figure
\ref{fig:mfout} shows the filter output $v'$ (equal to
$v'_{h}+v'_{n}$).  The presence of the ``signal'' $v'_{h}$ is now much
more evident.

The filter we have chosen has reduced the total power in the noise
relative to that in the signal. How it does this is apparent by
considering the power spectra in figures \ref{fig:mfin} and
\ref{fig:mfout}.  In figure \ref{fig:mfin}, the power in $v_{h}$ is
seen to be confined to a very narrow bandwidth about the frequency of
the damped sinusoid.  At its peak the signal power is about 5~dB
greater than noise power. Nevertheless, the total noise power,
integrated over the full bandwidth, is much greater than the signal
power and, consequently, the signal is overwhelmed by the noise (cf.\ 
the bottom panel of figure \ref{fig:mfin}).

Now consider $v'$.  The filter applied to the signal has the impulse
response of the signal, or the squared magnitude frequency response
given by the power spectrum in the top panel of figure \ref{fig:mfin}.
This is {\em matched\/} to the signal, in the sense that the power passed
is in the band where the signal power is large and the power stopped
is in the band where the signal power is small.  Thus, what survives
in $v'$ is the signal power, together with only that noise power in
the narrow band where the signal power is large. The signal to noise
of the filtered detector output $v'$ is correspondingly much higher in
the presence of the signal than is the signal to noise ratio of $v$. 

This example is illustrative.  In fact, we can ask, for an arbitrary
signal $v_{h}$ embedded in noise with power spectrum $S_{v}(f)$, for
the linear filter that maximizes the ratio of the mean-square signal
contribution to the mean-square noise contribution.  That filter is
referred to as the Wiener matched filter; in the frequency domain and
for weak signals it is (up to an overall constant)
\begin{equation}
    \widetilde{K}(f) = {\widetilde{v}_{h}^{*}(f)\over S_{h}(f)}.
\end{equation}

More generally, additional information is always useful for increasing
the our ability to detect a signal. This is true even that information
is not as complete as knowing the waveform. For example, consider the
case where we know the signal spectrum, but not its waveform. In the
frequency domain, we thus know the signal amplitude at each frequency,
but not the corresponding phase. In the case where the waveform is
known, we constructed the filter making full knowledge of both
amplitude and phase information. We can also construct a filter that
passes power in a given bandwidth, without regard to its phase. This
filter will emphasize power in the bands where the ratio of signal
power to noise power is relatively large over bands where the ratio is
small; consequently, it will increase our ability to detect a signal
whose spectrum is known in the same way that a matched filter
increases our ability to detect a signal whose waveform is known.

\subsection{The effective noise power spectral density}

How does one compare different detectors, with different response
functions and different noise power spectral densities?

One possibility is the ``performance benchmark'': choose a prototypical
source, evaluate the signal-to-noise that the source would give in the
different detectors, and determine finally which detector is most
likely to observe the source at a given level of confidence
\cite{finn97g}.

This kind of judgment depends critically on the source: using
different sources as your benchmark can lead to different conclusions.
For example, sources whose power is concentrated at different
frequencies focus attention on the detector noise at those
frequencies. Thus, while benchmarking detectors against particular
sources can be a powerful tool for comparing their relative
performance, it is also a tool with a very narrow focus. We need some
other way to compare the capability of detectors with a less specific
emphasis on source.

An important tool for making this more general comparison is the {\em
  effective power spectral density\/} $S_h(f)$,
\begin{equation}
    S_{h}(f) \equiv {S_{v}(f)\over|R(f)|^{2}},
\end{equation}
where $R(f)$ is the detector response function and $S_{v}(f)$ is the
detector noise power spectral density.  The quantity $S_{h}(f)$
describes an 
{\em effective detector noise:} it is the power spectral
density of a stochastic gravitational wave signal that would have to
be applied to a {\em noise-free\/} detector in order that the
corresponding response have power spectral density $S_{v}(f)$. Over
frequency bands where $S_{h}(f)$ is small, the detector is relatively
sensitive; over frequency bands where it is large, the detector is
relatively insensitive.

\begin{figure}
    \epsfxsize=0.8\columnwidth
    \begin{center}
        \leavevmode\epsffile{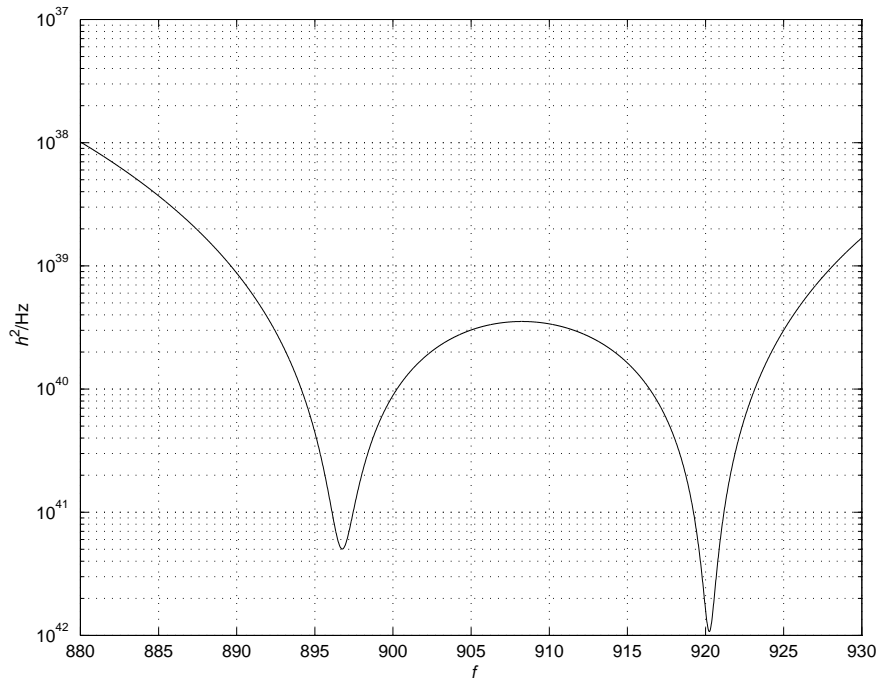}
    \end{center}
    \caption{The power spectral density of an effective stochastic 
    gravitational wave signal that would mimic the noise in the 
    output of a modern bar detector. Plotted is $\sqrt{S_{h}(f)}$ 
    {\em vs.} frequency $f$.}\label{fig:BarPSD}
\end{figure}

The effective power spectral density $S_{h}(f)$ has both a source and
detector independent meaning, making it a particularly useful quantity
for comparing gravitational wave detectors or for comparing a detector
to a source. With it, one can rank detectors according to their
overall noise in a given bandwidth, {\em e.g.,}
\begin{equation}
  \overline{h^2}_n(f_1,f_2) = \int_{f_1}^{f_2} df\, S_h(f), 
\end{equation}
or define an effective band $(f_0-\Delta f/2,f_0+\Delta f/2)$ over
which the detector has greatest sensitivity, {\em e.g.,}
\begin{eqnarray}
f_0 &\equiv& 
{\int_0^\infty df\, f/S_h(f)\over\int_0^\infty df/S_h(f)}\label{eq:f0}\\
\left(\Delta f\right)^2 &\equiv& 
{\int_0^\infty df\,(f-f_0)^2/S_h(f)\over\int_0^\infty df/S_h(f)}\label{eq:df}.
\end{eqnarray}
Finally, since the noise is referred directly to the amplitude of
incident gravitational radiation, one can calculate the expected SNR
of a given signal in the detector without reference to the detector's
response function:
\begin{eqnarray}
    \overline{\rho^{2}} 
    &=& 1 + 4\int_{0}^{\infty} df\,
    {\left|R(f)\widetilde{h}(f)\right|^{2}\over S_{v}(f)}\nonumber\\
    &=& 1 + 4\int_{0}^{\infty} df\,
    {\left|\widetilde{h}(f)\right|^{2}\over S_{h}(f)}.
\end{eqnarray}

\begin{figure}
    \epsfxsize=0.8\columnwidth
    \begin{center}
        \leavevmode\epsffile{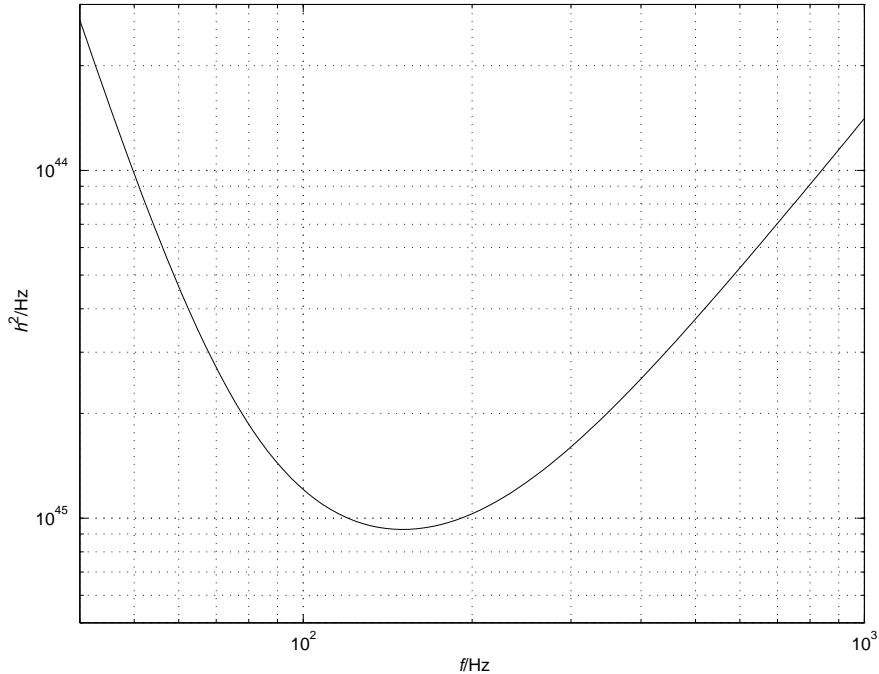}
    \end{center}
    \caption{The power spectral density of an effective stochastic 
      gravitational wave signal that would mimic the noise in the
      output of first generation LIGO instrumentation. Plotted is
      $\sqrt{S_{h}(f)}$ {\em vs.} frequency $f$.}\label{fig:IFOPSD}
\end{figure}

Figure \ref{fig:BarPSD} shows the modeled $S_{h}(f)$ for a modern bar
detector, while figure \ref{fig:IFOPSD} shows $S_{h}(f)$ for a model
of the first-generation LIGO instrumentation.  Note how the bar
detector noise is particularly small in two narrow
bands\footnote{Since the bar detectors ``sensitivity'' $1/S_{h}$ is
  multi-modal it is more appropriate to define the effective band, as
  in eq.\ \ref{eq:f0} and \ref{eq:df}, separately
  about each peak.} about the resonant frequencies of the two mode
system consisting of the bar and its transducer, while the
interferometer achieves its peak sensitivity over a much broader
bandwidth.

\subsubsection{An aside: noise in bar detectors}

It is a common misconception that bar detectors are intrinsically
narrow-band detectors. While the amplitude of a resonant detector's
response is greatest for signal power in the neighborhood of the
resonance, the thermal excitation of the bar is also concentrated in
this band as well. The net result is that the contribution of the
bar's thermal noise to the power spectral density expressed in units
of $h^2/\mbox{Hz}$ is effectively independent of frequency.

To understand how resonant detectors become narrow band instruments,
consider how the signal appears in the electronics that follow the
transducer.  The resonant character of the detector leads to large
amplitude motion for signal power near the resonant frequency and
small amplitude motion for signal power far from the resonance. 
Correspondingly, the amplified signal is large near to, and small far
from, the resonance.  {\em The amplifier contributes its own noise,
however, which is approximately white at the amplifier output.} Thus,
compared to the signal presented for amplification, the amplifier
noise is {\em relatively\/} large far from resonance and {\em
relatively\/} small near to resonance.

In present-day resonant cryogenic detectors the bandwidth is limited
by amplifier noise, referred back to $h$ through the response
function.  

Since it is the amplifier noise, when referred back to $h$ through the
response function of the resonant bar, that limits the instrument
bandwidth, why make the bar resonant at all? The purpose of making the
detector resonant is to provide {\em mechanical\/} amplification for the
signal, so that, at least in a narrow bandwidth, it is much stronger
then the limiting noise source: amplifier noise.  Signal power at or
near resonance leads to a large excitation of the bar, which
translates into a large input to the amplifier; thus, the resonance of
the bar amplifies the incident gravitational wave signal relative to
all the noise sources that follow, including the limiting amplifier
noise. 

\subsection{Conclusion}

Gravitational wave detectors are characterized by their antenna
patterns, which describe their differential sensitivity to radiation
incident from different directions and with different polarizations,
their response functions, which describe the differential amplitude of
their response to signals of different frequency, and the character of
their noise. 

The distinction between the response function and the antenna pattern
is sometimes an artificial one: the response function can (and for
interferometric detectors, does) depend on the incident direction of
the radiation. 

Much of the experimental craft is devoted to making the detector noise
approximately stationary and Gaussian (or in making the signal so
large that the character of the noise is not significant for
measurements of interesting precision). Stationary noise can be
characterized by evaluating the correlations among samples taken at
different relative times. For Gaussian noise, all the correlations are
known once the pairwise correlation function is measured. 

While the correlation functions are good conceptual tools for
understanding the character of stationary detector noise, a more
useful tool, fully equivalent, is the noise spectrum: the Fourier
transform of the correlation function on its time arguments. The noise
power spectral density, in particular, is a particularly important and
useful tool for characterizing the disposition of detector noise
power. 

\section{Characterizing detection}\label{sec:charToDet}

What does ``detection'' mean? Let's try to frame an answer by posing a
specific question --- {\em e.g.,} ``with what confidence can we
conclude that, in the last hour, gravitational waves from a new
core collapse supernova in the Virgo cluster of galaxies passed
through our gravitational wave detector?'' --- and exploring its
meaning.

It turns out that, straightforward as it seems, there are two
different ways of interpreting this question; correspondingly, there
are two different meanings that ``detection'' can take. How we mean
the question --- or the kind of answer that we want to take away ---
determines the kind of analysis that we need to undertake with data
collected at a gravitational wave detector. 

\subsection{Learning From Observation}\label{sec:Bayesian} 

``With what confidence can we conclude that, in the last hour, 
gravitational waves from a new core collapse supernova in the Virgo
cluster of galaxies passed through our gravitational wave detector?''

As with most questions of detection, even before examining our
observations we have some expectation of the answer. In this case, we
know the rate of supernovae and this leads us to expect, on average,
one such core collapse every 4 months; consequently, we believe the
probability is approximately $3.4\times10^{-4}$ that in any given hour
--- including the last --- gravitational waves from a new Virgo
cluster supernova were incident on our detector.

Probability, as we have used it here, means {\em degree of belief.} In
this instance, our degree of belief coincided with the {\em expected
  frequency\/} of supernova events. This need not always be the case:
we can assess degree of belief even when we can't assess relative
frequency.  For example, suppose that I have a coin that is known to
be heavily biased toward either heads or tails.  What is your degree
of belief that, when I next flip the coin, it will land heads-up?
Without telling you the direction or amount of the bias, you can't
evaluate the expected relative frequency of heads or tails.  You can,
however, quantify your degree of belief: having no more reason to
believe that the bias is toward heads than towards tails, you have no
more reason to believe that the coin will, when next flipped, land
heads-up than that it will land heads-down.  Your {\em degree of
  belief\/} in either alternative, then, is $1/2$.

One does not have to search either long or hard to find examples from,
{\em e.g.,}
astrophysics, where probability as ``degree of belief'' exists and
probability as ``expected frequency'' does not.  For example, what is
the probability that there exists a cosmological stochastic
gravitational wave signal with a given amplitude and spectrum?  In this
case, ``expected frequency'' has no meaning: there is only one
Universe, and it either does or does not have a stochastic
gravitational wave background of given spectrum and amplitude.

After we examine the output of our gravitational-wave detector, our
degree of belief in the supernova proposition may change: we may, on
the basis of the observations, become more or less certain that
radiation from a supernova passed through our detector.  How do
observations change our degree of belief in the different
alternatives?

To explore how our degree of belief evolves with the examination of 
observations we need to introduce some notation:
\begin{eqnarray}
H_{0} &=& \left(\begin{array}{l}
\mbox{proposition that gravitational waves from a}\\
\mbox{new supernova in the Virgo cluster {\em did not}}\\
\mbox{pass through our detector in the last hour}
\end{array}
\right),\\
{\cal I}&=& \left(\begin{array}{l}
\mbox{our prior knowledge of astrophysics, including}\\
\mbox{our best assessment of the supernova rate}
\end{array}\right),\\
g &=& \left(\begin{array}{l}
\mbox{observations from our gravitational wave detector}
\end{array}\right),\\
P(A|B) &=& \left(\begin{array}{l}
    \mbox{degree of belief in $A$ assuming that $B$ is true}
\end{array}\right),\\
\neg{}A &=& \left(\mbox{logical negation of proposition $A$}\right).
\end{eqnarray}
In this notation, $P(H_{0}|{\cal I})$ is the degree of belief we
ascribe to the proposition that no gravitational waves from a core
collapse supernova in the Virgo cluster passed through our detector in
the last hour, given only our prior understanding of astrophysics;
similarly, $P(H_{0}|g,{\cal I})$ is the degree of belief we ascribe to
the same proposition, give {\em both} the observation $g$ and our
prior understanding of astrophysics.

To understand how $P(H_{0}|{\cal I})$ and $P(H_{0}|g,{\cal I})$ are
related to each other we need to recall two properties of probability.
The first is unitarity: probability summed over all alternatives is
equal to one. In our example, the two alternatives are that
a supernova occurred or it did not:
\begin{equation}
P(H_{0}|g,{\cal I}) + P(\neg{}H_{0}|g,{\cal I}) = 1.
\end{equation}
The second property we need to recall is Bayes Law, which describes
how conditional probabilities
% ({\em e.g.,} $P(A|B,C)$, which is the 
% probability of $A$ given that both $B$ and
% $C$ are true) 
``factor'':
\begin{equation}
P(A|B,C)P(B|C) = P(A,B|C) = P(B|A,C)P(A|C).
\end{equation}
Combining unitarity and Bayes Law it is straightforward to show that
\begin{equation}
P(\neg{}H_{0}|g,{\cal I}) = {\Lambda(g)\over\Lambda(g) +
P(H_{0}|{\cal I})/P(\neg{}H_{0}|{\cal I})}
\label{eq:p(H0|g)}
\end{equation}
where
\begin{eqnarray}
\Lambda(g) &=& {P(g|\neg{}H_{0},{\cal I})/ P(g|{H}_{0},{\cal I})}\\
P(g|H_{0},{\cal I}) &=& \left(\begin{array}{l}
\mbox{probability that $g$ is a sample of}\\
\mbox{detector output when $H_{0}$ is true} 
\end{array}\right)\label{eq:p(g|H0)}\\
P(g|\neg{}H_{0},{\cal I}) &=& \left(\begin{array}{l}
\mbox{probability that $g$ is a sample of}\\
\mbox{detector output when $H_{0}$ is false} 
\end{array}\right)\label{eq:p(g|notH0)}
\end{eqnarray}

The two probabilities $P(g|H_{0},{\cal I})$ and $P(g|\neg{}H_{0},{\cal
I})$ depend on the statistical properties of the detector noise and
the detector response to the gravitational wave signal.  In some cases
they can be calculated analytically; in other circumstances it may be
necessary to evaluate them using, {\em e.g.,} Monte Carlo numerical
methods.  Regardless of how one approaches data analysis the detector
must be sufficiently well characterized that these or equivalent
quantities are calculable.

Equation \ref{eq:p(H0|g)} describes how our degree of belief in the
proposition $\neg{}H_{0}$ evolves as we review the observations.  If
$\Lambda$ is large compared to the ratio $P({H}_{0}|{\cal
  I})$ to $P(\neg{}H_{0}|{\cal I})$ then our confidence in $\neg{}H_{0}$
increases; alternatively, if it is small, then our confidence in
$\neg{}H_{0}$ decreases.  If $\Lambda$ is equal to unity --- {\em
  i.e.,} the observation $g$ is equally likely given $H_{0}$ or
$\neg{}H_{0}$ --- then the posterior probability $P(H_0|g,{\cal I})$
is equal to the prior probability $P(H_0|{\cal I})$ and our degree of
belief in $H_0$ is unchanged: we learn nothing from the observation.

We can now answer the question that began this section. 
We understand confidence to mean {\em degree of belief\/} in the
proposition that radiation originating from a new supernova in the
Virgo cluster was incident on a particular detector during a
particular hour. In response we make a quantitative assessment of
our degree of belief in that proposition --- the probability that the
proposition is true.

\subsection{Guessing Natures State}\label{sec:Frequentist} 

Begin again: ``With what confidence can we conclude that, in the last
hour, gravitational waves from a new core collapse supernova in
the Virgo cluster of galaxies passed through our gravitational wave
detector?''

As before, we have the hypothesis $H_{0}$ and its logical negation, 
$\neg{}H_{0}$.  The gravitational waves from a new Virgo cluster 
supernova either passed through our detector, or they did not.  Our 
goal is to determine, as best we can, which of these two alternatives 
correctly describes what happened.

We decide which alternative is correct by consulting our observation
$g$.  Operationally, we adopt a rule or a procedure that, when applied
to $g$, leads us to accept or reject $H_{0}$.  The question that began
this section asks us for our degree of confidence in the most reliable
rule or procedure.

There are many procedures that we can choose from.  Some are just
plain silly: for example, always rejecting $H_{0}$ is a procedure. 
Similarly, accepting $H_{0}$ if a flipped coin lands heads is a
procedure.  Some procedures are more sensible: we can calculate a
characteristic amplitude from the observation ({\em e.g.,} a
signal-to-noise ratio) and reject $H_{0}$ if the amplitude exceeds a
threshold.  Nature doesn't always speak clearly; additionally, some
crucial information is often hidden from us.  Consequently, no
procedure will, in the end, be perfect and every rule will, on
unpredictable occasions, lead us to erroneous conclusions.  Still,
some procedures are clearly better than others: the question is, how
do we distinguish between them quantitatively?

Better procedures are those that are less prone to error. 
Consequently, we focus on the frequency with which different
procedures err.  For our simple problem, where we want to decide only
if we have or have not observed the radiation from a supernova (reject
or accept $H_{0}$), there are two kinds of errors a decision procedure
can make:
\begin{enumerate}
\item If no radiation is present ($H_{0}$ true), the rule may
  incorrectly lead us to conclude that radiation is present: 
  a {\em false alarm,} or type I, error.
\item If radiation is present ($H_{0}$ false), the rule may
  incorrectly lead us to conclude that radiation is absent: 
  a {\em false dismissal,} or type II, error.
\end{enumerate}
The false alarm frequency is generally denoted $\alpha$ while the 
false dismissal frequency is denoted $\beta$.

If we have an ensemble of identical detectors, each observing
simultaneously the same system for which $H_{0}$ (or $\neg{}H_{0}$) is
true, and we apply our rule to each observation, then the fraction of
observations in the ensemble that lead us to reject (accept) $H_0$ is
just the false alarm (dismissal) frequency.  False alarm and false
dismissal frequencies can be interpreted as probabilities: in
particular, the probability of our rule giving an incorrect result.

Even in the simple case at hand (a single hypothesis that we must
accept or reject), there are at least two distinct kinds of errors
that an inference procedure can make.  Our measure of a rule's
reliability thus involves at least two dimensions, and may involve
more.  How, then, do we order rules to settle upon a best, or optimal,
rule?

To rank rules we must reduce the several error measures that describe 
a procedure's performance to a single figure of merit.  How we choose 
to do this depends on the nature of our problem.  In our case, rules 
that distinguish between $H_{0}$ and $\neg{}H_{0}$ are characterized 
by their false alarm and false dismissal frequencies; consequently, our 
criteria for ranking rules should depend on our relative intolerance to 
false alarms and false dismissals.  For example, if we are testing for 
the presence of antibodies in an effort to diagnose and treat a 
serious illness, we might be very concerned to keep the false 
dismissal rate low, and not nearly as worried about a high false alarm 
rate: after all, a false dismissal might result in death, while a 
false alarm only in an unnecessary treatment with less serious 
repercussions.  Judges or juries in criminal trials faces different 
concerns: false dismissals let criminals go free, while
false alarms send the innocent to prison --- neither alternative 
being very palatable.  Finally, in the case of gravitational wave 
detection, we may (at least initially) be very concerned to avoid 
false alarms, even at the risk of falsely dismissing many real 
signals.

Thus, in order to provide a relative ranking of different inference
procedures for detection or parameter estimation we must construct an
{\em ad hoc\/} figure of merit that reflects our sensitivity to an
incorrect decision.  We term the best rule, under that {\em ad hoc\/}
criteria, the ``optimal'' rule.  ``Optimality'', however, is a
relative concept: if the criteria change, the ``optimal'' rule changes
also.  In the three examples given above, the criteria might be
\begin{itemize}
\item {\em medical diagnosis:} fix a maximum acceptable false
  dismissal rate and choose the rule that, among all rules whose false
  dismissal rate is so constrained, has the minimum false alarm rate;
\item {\em criminal justice:} choose a rule whose weighted total error
  $\alpha\cos\phi+\beta\sin\phi$ is minimized ($\phi$ being a matter
  of personal choice for an individual judge or juror); 
\item {\em gravitational wave detection:} fix a maximum acceptable
  false alarm rate and choose the rule that, among all rules whose
  false alarm rate is so constrained, has the minimum false dismissal
  rate. 
\end{itemize}

False alarm and dismissal rates describe our confidence in the 
long-run behavior of the associated decision rule.  To understand the 
implications of this measure of confidence, suppose that we have not 
one, but $N$ independent and identical detectors all observing during 
the same hour.  We use the same test, with false alarm rate $\alpha$ 
and false dismissal rate $\beta$, on the observations made at each 
detector, and find that, of these $N$ observations, $m$ lead us 
(through our inference rule) to reject $H_0$ and $N-m$ lead us to 
accept $H_0$.  For a concrete example, suppose $\alpha$ is 1\%,
$N$ is ten and $m$ is three.

The probability of obtaining this outcome when the signal is absent
($H_0$ is true) is the probability of obtaining $m$ false alarms in
$N$ trials, or
\begin{equation}
P(m|H_0,N) = {N!\over(N-m)!m!}\alpha^m(1-\alpha)^{N-m}.
\end{equation}
In our example, $P(m|H_0,N)$ evaluates to $1.1\times10^{-4}$. It is
thus very unlikely that we would have made this observation if the
signal were absent. Does this mean we should conclude the signal is
present with, say, $99.99$\% confidence?

{\em No!} $P(m|H_0,N)$ describes the probability of observing $m$ false
alarms out of $N$ observations.  When the signal is {\em present,}
however ({\em i.e.,} when $H_0$ is false), there are {\em no\/} false
alarms and both $\alpha$ and $P(m|H_0,N)$ are irrelevant. There are,
however, $N-m$ false dismissals; thus, the relevant quantity is
$P(m|\neg{}H_0,N)$, the probability of observing $N-m$ false {\em
  dismissals:}
\begin{equation}
P(m|\neg{}H_0,N) = {N!\over(N-m)!m!}(1-\beta)^m\beta^{N-m}.
\end{equation}
If, in our example, the false dismissal rate $\beta$ is 10\%, then the
probability of observing seven false dismissals out of ten trials is
is $8.7\times10^{-5}$.

The particular outcome of our example --- three positive results out
of ten trials --- is, in the grand scheme of things, very unlikely;
nevertheless, what is important to us is that it is {\em more\/}
unlikely to have occurred when the signal is present then when it is
absent.  Despite the apparently overwhelming improbability of three
false alarms in ten trials, it is nevertheless, slightly more likely
than the alternative of seven false dismissals in ten trials.

We can now answer the question that began this section.  We understand
that question to ask for the error rate of the best general procedure
for deciding between the alternative hypotheses.  There is an implicit
assumption regarding the decision criteria, which tells us what
``best'' means in this context.  In the context of these criteria, we
calculate the error rates for different inference rules, choose rank
the different rules, and find the best rule and its corresponding
error rates.

Contrast this with our understanding of the identically worded
question as we understood it in the previous subsection. There, we
understood confidence to mean the degree of belief that we should
ascribe to alternative hypotheses; here, we understand confidence to
refer to the overall reliability of our inference procedure. There we
responded with a quantitative assessment of our degree of belief in
the alternative hypotheses, {\em given a particular observation made
  in a particular detector over a particular period of time;} here we
responded with an assessment of the relative frequency with which our
rule errs given each alternative hypothesis. There we did not make a
choice between alternative hypotheses; rather, we rated them as more
or less likely to be true in the face of a particular observation.
Here, on the other hand, we do make choices and our concern is with
the error rate of our procedure for choosing, averaged over many
different observations and many different decisions.

Analyses like the ones in this section, where probability is interpreted
as the limiting frequency of repeatable events and the focus is on
false alarm and false dismissal frequencies, are termed Frequentist
analyses. They have particular utility when it is possible to make
repeated observations on identical systems: {\em e.g.,} particle
collisions in an accelerator, where each interaction of particle
bunches is a separate ``experiment.'' Analyses like those in the previous
section, where probability is interpreted as degree-of-belief and
the focus is on the probability of different hypotheses conditioned on
the observed data, are termed Bayesian analyses. Bayesian analyses are
particularly appropriate when the observations or experiments are
non-repeatable: {\em e.g.,} when the sources are, like supernovae,
non-identical and destroy themselves in the process of creating the
signal. In this case we are interested in the properties of the
individual systems and would prefer a measure of the relative degree
of belief that we should ascribe to, for example, the proposition that
the signal originated from a particular point in the sky.

That Bayesian and Frequentist analyses are different does not imply
that one is right and the other wrong.  Bayesian and Frequentist
analyses do not address the same questions; so, they are not required
to reach ``identical'' conclusions.  On the other hand, it may well be
that one analysis is more appropriate or responsive to our concerns
than the other.  We can only make the choice of appropriate analysis
tools when we understand the distinction between them.

\section{Gravitational Radiation Sources}

In this section we review briefly some of the different kinds of
sources that are, at this writing, thought to be ``important'' for the
generation of large interferometric detectors now under construction.
``Important'' is a term that requires definition in this context.  Clearly,
sources that we don't expect to detect, or to be able to detect, are
unimportant. Detectable sources must radiate significant energy in the
bandwidth where these detectors are most sensitive (ranging from the
tens to hundreds of Hz) over a reasonable observation period. For
periodic sources, this is the integrated power over a period of
several months to perhaps as much as one to two years; for burst
sources, this means that the expected rate of detectable bursts must
be at least several per year. 

The assessment of source strength and number or rate is difficult to
make for most sources. Very often the radiation strength
depends on physics and astrophysics that we don't know or understand
in the requisite detail. For all proposed sources the rate, number or
even existence of sources whose signal strengths are large enough to
be detectable is difficult to ascertain. This is not surprising: what
we know of the heavens we know principally through electromagnetic
observations; however, it is in the nature of gravitational wave sources
that they leave little electromagnetic evidence of their existence.

Finally, since we are, with these instruments, looking at the universe
in a fundamentally new way, we must keep an open mind to the
possibility of sources unimagined: in this I side with John Haldane,
who said (in a different context) ``My own suspicion is that the
universe is not only stranger than we suppose, but stranger than we
can suppose.''

One final note: until now, we have been careful to keep all factors
$G$ and $c$ in our expressions for gravitational effects. Here and
henceforth, we will write all expressions in units where $G$ and $c$
are unity: {\em e.g.,} units of length, with conversion factors from
grams and ergs to centimeters as given in equations
\ref{eq:units-mass} and \ref{eq:units-energy}. These conversion
factors can be invoked to find expressions in terms of quantities
expressed in more conventional units.

\subsection{Burst sources}\label{sec:bursts}

\subsubsection{Compact binary inspiral}

The source most-talked-about for the interferometric detectors now
under construction are binary systems consisting of two compact,
stellar mass objects --- either neutron stars or stellar mass black
holes.  Like a rotating dumbbell, a binary star system has a large,
accelerating quadrupole moment, which makes it (for its mass) a strong
gravitational radiation source.  The radiation carries away orbital
binding energy and orbital angular momentum, which leads to a faster
and more compact orbit.  Kepler's Third Law relates the orbital
frequency $f_{\mbox{\small orb}}$, semi-major axis $a$ and total
system mass $M$ by
\begin{equation}
f_{\mbox{\small orb}}^2 = {M\over 4\pi^2 a^{3}}.
\end{equation}
Consequently, the radiated power is, in order of magnitude,
\begin{equation}
L \propto \left[Ma^2f_{\mbox{\small orb}}^3\right]^2. 
\end{equation}
Note that radiated power {\em increases\/} as the orbit decays:
\begin{equation}
L \propto \left(M\over a\right)^{5}.
\end{equation}
For binaries that can become sufficiently compact the power radiated
gravitationally will, in the end, become large enough to dominate the
system's evolution.  Since the radiated power increases as the orbit
decays, the system will then decay at an ever increasing rate, with
ever increasing radiation amplitude, frequency and power, until the
components coalesce.  It is the radiation from this {\em inspiral,}
for binary systems of neutron stars or black holes, that is seen as an
important source for the LIGO and VIRGO detectors.

Why compact components, like neutron stars or black holes, and why
stellar mass, and not more or less massive? Recall that the proposed
interferometric detectors have their greatest sensitivity at 
approximately 150~Hz. The quadrupole radiation from a
binary system is at twice the system's orbital frequency;
correspondingly, if the radiation is to be in the bandwidth of these
detectors, the binary systems themselves must exist with orbital
frequencies of at least 75~Hz. Kepler's Third Law places a lower bound
for us on the matter density of the components, which must
be much greater than the total system mass divided by the cube of the
orbital radius:
\begin{eqnarray}
\rho_{1,2} &\gg& {M\over a^3}\\
&\simeq& \left(\pi f_{\mbox{\small gw}}\right)^{2}\\
&\simeq& 1.5\times10^{12} \left(f_{\mbox{\small gw}}\over
  100\,\mbox{Hz}\right)^2 \mbox{g/cm$^3$}.
\end{eqnarray}
Thus, irrespective of the total system mass, if a binary system is to
radiate in a band where these detectors are sensitive the central
density of its components cannot be much less than nuclear density.
With this we are forced, for astrophysical objects, to restrict
attention to neutron stars or black holes.

The nuclear and super-nuclear equation of state place upper limits on
the neutron star mass, which does not apply for a black hole.  The
dynamics of the binary orbit, however, does place an upper limit on
the mass of the black hole binaries that the ground-based
interferometric detectors may observe.  With every orbit the binary
radiates away more of its binding energy, leading to a more compact
orbit.  Eventually the system coalesces: the two components merge,
collide, or tidally disrupt.  Even if we imagine that the components
are point masses, so that there is no tidal disruption or collision
that would terminate the inspiral signal at some finite orbital
frequency, relativity appears to impose a maximum orbital frequency on
binary systems.  For approximately symmetric binary star systems ({\em
i.e.,} those with equal mass components) this limit is\cite{kidder92a}
\begin{equation}
 f_{\max} \simeq 710 {2.8\,\mbox{M}_{\odot}\over M} \,
 \mbox{Hz},
 \label{eq:kww}
\end{equation}
where $M$ is the system's total mass.  Thus, the component black hole
masses must be less than 15~$\mbox{M}_{\odot}$ if the inspiral signal
is to survive into the bandwidth where the detector is most sensitive.

It is currently thought that, during the epoch when the radiation from
the binary is in the bandwidth where the LIGO and VIRGO detector
sensitivity is greatest, the binary components are well approximated as
point masses for the purpose of computing the radiation and orbital
evolution\cite{bildsten92a} (There is some small suggestion that
resonant tidal interactions may complicate this picture\cite{ho98a}.) 
During this epoch, the gravitational fields that determine the binary
evolution are sufficiently strong that first order perturbation theory
is not adequate to compute the orbits; nevertheless, the fields are
not so strong that computing the orbits and the radiation via higher
order perturbation theory is impractical
\cite{blanchet98a,blanchet95a}.  For this overview, no additional
insight is gained by considering anything higher the quadrupole
formula radiation, in which case the excitation of the detector --- an
effective $h(t)$ that is a superposition of the radiation in the two
polarization states of the wave --- is\cite{finn93a,finn96a}
\begin{equation}
h(t) = {{\cal M}\over d_L}\Theta\left(\pi f{\cal
    M}\right)^{2/3}\cos\Phi(t),
\end{equation}
where
\begin{eqnarray}
{\cal M} &\equiv&
{\left(m_1m_2\right)^{3/5}\over\left(m_1+m_2\right)^{1/5}}(1+z),\\
\Theta^2 &\equiv& 4\left[
F_{+}^2(1+\cos^2\iota)^2 + 4F_{\times}^2\cos^2\iota
\right],\\
f(t) &\equiv& {1\over\pi{\cal M}}
\left({5\over256}{{\cal M}\over T_0-t}\right)^{3/8},\label{eq:fscale}\\
\Phi(t) &=& \int^t 2\pi f(t')dt',
\end{eqnarray}
$d_L$ is the cosmological luminosity distance to the source, $m_1$ and
$m_2$ are the binary system's component masses, $z$ is the source's
cosmological redshift, $\iota$ is the angle between the binary's
angular momentum axis and the line of sight to the detector, and $T_0$
is a constant of integration.

What can we determine through observation of the signal from such a
system? The signal-to-noise, of course, which takes on a particularly
simple form\cite{finn93a,finn96a}:
\begin{equation}
\overline{\rho^2} \simeq 1 + 25 \left({r_0\over d_L}\right)^2 
\left({\cal M}\over
  1.2\,\mbox{M}_{\odot}\right)^{5/3}, 
\end{equation}
where $r_0$ is a characteristic distance that depends only on the
effective power spectral density of the source, 
\begin{equation}
r_0^2 = 
\left(G\mbox{M}_{\odot}\over c^2\right)^2 
{5\over 192\pi}  
\left(243\over7\times10^5\right)^{1/3}
\int_0^\infty {
df\,\left(\pi G\mbox{M}_{\odot}/c^3\right)^2 \over
\left(\pi fG\mbox{M}_{\odot}/c^3\right)^{7/3}S_h(f)
}\label{eq:r0}
\end{equation}
and the average denoted by the over-bar is over both an ensemble of
detectors and all relative orientations of the source and the
detector.  For the initial LIGO and VIRGO detectors, $r_0$ is about
13~Mpc.  In order that we are confident that we have seen a source,
the SNR $\rho^{2}$ should not be much less than about 65 in a single
detector; so, we don't expect to see sources from distances beyond
more than a few $r_0$.

How many of these sources can LIGO expect to see? Unfortunately, we
know very little about the rate of compact binary coalescence, except
that it is rare. Black-hole/black-hole binaries are, of course,
invisible to us except through gravitational waves. Binaries involving
neutron star component(s) are observable to us only if one of the
components is a pulsar. Pulsars are observable only if they are not
too distant (in our galaxy or its satellite globular clusters) and if
the pulsar beam intersects our line of sight. 

There are only three known binary pulsar systems in or about our own
galaxy that will coalesce in less than the age of the universe; of
these three, one in particular drives the calculation of the rate
density.  If we attempt to project this meager observational data
throughout the entire universe, accounting for observational biases
that cause us to miss some fraction of the actual number of binary
systems, we find that the rate density of coalescing binaries is
\begin{equation}
\dot{n} \simeq 10^{-7\pm2}\,\mbox{Mpc}^{-3}\,\mbox{yr}^{-1}.
\end{equation}
The great uncertainty in how to project the observations throughout
the universe is reflected in the factor of $10^4$ uncertainty in this
rate. (For a sense of the uncertainties and corresponding controversy
surrounding the estimates of the rate density of inspiraling binary
coalescence, see
\cite{stairs98a,bethe98a,bethe98b,bagot98a,yungelson98a,narayan91a,phinney91a}
and references therein.)

The quantity $r_0$, defined in equation \ref{eq:r0}, was constructed
in such a way that, assuming that sources are distributed
homogeneously and isotropically throughout space, the rate of sources
observed above a signal-to-noise $\rho$ of $8$ ($\rho^2$ of 64) is
equal to
\begin{equation}
\dot{N} = {4\pi\over3}r_0^3\dot{n}. 
\end{equation}
For the initial LIGO detectors, $r_0$ is only about $13\,\mbox{Mpc}$;
so, the {\em anticipated\/} rate of binary inspiral is, even at its
most optimistic, low for the first generation of detectors.  Things
get better for the more advanced instruments now on the drawing board:
for these, $r_0$ climbs to over $100\,\mbox{Mpc}$ and, by
correlating the signal from the two LIGO 4~Km detectors and the
LIGO 2~Km detector, the effective $r_0$ can be increased by another factor
of 3/2.\cite{finn96a}

In addition to the SNR, we also observe the scaling of the radiation
frequency with time. From equation \ref{eq:fscale} this gives us
${\cal M}$, the so-called chirp mass, which depends on the component
masses and the source's cosmological redshift.  Knowing both the
SNR and the chirp mass raises the interesting possibility of measuring
the Hubble constant: the rate of cosmological expansion. Measuring the
SNR tells us, up to the complications of the noise and the unknown
orientation angles, something about the luminosity distance to the
source.  Similarly, measuring the chirp mass tells us, up to the
unknown component masses, something about the source redshift. There
is thus a hidden redshift/luminosity-distance relationship in
observations of binary inspiral. By statistical analysis of a large
number of binary inspiral observations, that relationship can be
extracted and with it the Hubble constant\cite{finn96a}.

\subsubsection{Compact binary coalescence}

Eventually the inspiraling orbit of a binary system with compact
components must end: the neutron stars collide, the black holes merge,
or the one black hole tidally disrupts its neutron star companion. 
The radiation arising from the last few inspiral orbits through the
coalescence of the compact objects that compose the binary may also be
a significant source of radiation.

Unfortunately, very little is really known about the gravitational
waves that result from the late stages of inspiral or the coalescence
of either neutron star or black hole binary systems.  In both cases
the gravitational fields are quite strong and dynamical, which would
appear to rule-out a perturbative approach and require a (numerical)
solution to the fully non-linear Einstein field
equations.\cite{finn97c}  (In the neutron star case, the problem is
further complicated by the need to model the dynamics of the fluid,
which cannot be ignored in a coalescence.)

Numerical simulations of the coalescence of two black holes in a
head-on collision have been calculated and provide some guidance:
these tend to show that the total energy energy is disappointingly
small: on order $10^{-4}$ of the system's total mass-energy
\cite{smarr79b,anninos93a,anninos95d,baker97a,anninos98a}.  There are
also some recent calculations of off-axis collisions, which suggest
strongly that the maximum energy radiated in such a collision will be
less than 1\% of the total mass energy\cite{khanna99a}. In particular,
it is difficult to justify the additional, {\em ad hoc\/} factor of
10\%$\mbox{M}c^2$ assumed by some authors in estimating the
detectability of these sources.\cite{flanagan98a,flanagan98b}.

Surprisingly, perturbation calculations of the radiation arising from
colliding black hole spacetimes give results that are in close accord
with the limited number of fully relativistic numerical simulations
that have been
performed\cite{price94a,anninos95e,abrahams96d,abrahams96e,price94a,andradea97a,baker97a,gleiser98a}.
 This accord is difficult to explain and may signal that we have
something new to learn about the nature of the solutions to the full
field equations.

\subsubsection{Black Hole Formation}

Black holes form in the collision of neutron stars at the end-point of
neutron star binary inspiral; they also form in the core collapse of
sufficiently massive stars. Unless the formation mechanism is
especially symmetric, the new black holes that form will be initially
quite deformed and will need to radiate away their deformations before they
can settle down into a quiescent state, which is axisymmetric.

Quiescent black holes are characterized only by their mass $M$ and
angular momentum $J$.  (And electric charge, too; however,
astrophysical black holes are unlikely to carry any significant
electric charge.)  Correspondingly, while the initial radiation from
the formation of a black hole depends on the details of the formation,
the final radiation depends principally on $M$ and $J$.  In fact, the
late-time waveform from a perturbed black hole is a superposition of
exponentially damped sinusoids, whose frequencies and damping times
depend only on $M$ and $J$, the overtone number $n$ and the harmonic
order $\ell$ and $m$ of the perturbation.

Most all of the modes of a black hole are very strongly damped. 
The most weakly damped modes are
associated with the fundamental quadrupole-order excitations.  Even
these are strongly damped unless the black hole is very rapidly
rotating.  For this reason, we focus attention on the fundamental
quadrupole modes, which are the most likely to be detectable.  Setting
aside the start-up transient associated with the details of the
initial excitation, a good model for the ``ring-down'' of a
newly-formed or perturbed black hole is
thus\cite{echeverria89a,finn92a}
\begin{equation}
h_{\mbox{\tiny RMS}}(t) = 2\sqrt{2\epsilon\over Q(a)F(a)}{M\over r}
e^{-\pi ft/Q}\sin\left(2\pi ft\right) \qquad\mbox{($t>0$)},
\end{equation}
where the amplitude is averaged (in a root-mean-square sense) over all
orientation angles,
\begin{eqnarray}
f&\simeq&12.\,{\rm KHz}\left({\mbox{M}_\odot\over
    M}\right)\left(F(a)\over 37/100\right),\\
Q&\simeq&2(1-a)^{-9/20},\\
a &\equiv& {J\over M^2},\\
F(a) &\simeq& 1-{63\over100}(1-a)^{3/10},
\end{eqnarray}
$r$ is the distance from the black hole to the detector, $\epsilon$
is the fraction of the total mass of the black hole carried away in
radiation, and we have assumed that all five of the fundamental tone
quadrupole modes are excited equally. Corresponding to this radiation
is an estimated signal-to-noise ratio of
\begin{equation}
\overline{\rho^2} = 1+
34\,G(a)^2{\epsilon\over 10^{-4}}
\left(20\,\mbox{Mpc}\over r\right)^2
\left(M\over 13\,\mbox{M}_\odot\right)^3
{10^{-46}\,\mbox{Hz}^{-1}\over S_h},
\end{equation}
where we have assumed 
\begin{itemize}
\item an efficiency $\epsilon$ for fraction of the rest mass of the
  system radiated gravitationally that is equivalent to what is found
  in black hole collisions, and
\item the effective noise power spectral density is approximately
  constant over the signal bandwidth (which is broad for strongly
  damped oscillations).
\end{itemize}

The rate of black hole formation is entirely uncertain; however, most
astrophysicists see no reason why the same mechanisms that make
neutron stars cannot also make black holes at approximately the same
rate.\cite{narayan91a} By our present understanding of formation
mechanisms, this rate is not high even at the distance of the Virgo
cluster ($\sim20\,\mbox{Mpc}$): perhaps as many as a few per year, but
likely much less.  Consequently, $\overline{\rho^2}$ should be at
least on order 30--35 for a confident detection in ideal
circumstances.\cite{finn91a} The caveat of ``ideal circumstances'' is
an important one, however: the character of the waveform for this
source --- an exponentially damped sinusoid --- is exactly the kind of
technical noise one might expect in a real interferometer owing to
transient disturbances that affect, for example, the suspension of the
interferometer mirrors.  Thus, without strong assurance that what is
observed is not a weak disturbance intrinsic to the detector,
prospects are not good for observing radiation from this source.

\subsubsection{Stellar Core Collapse}

Theoretical models of stellar core collapse, and the corresponding
gravitational wave luminosity, have a long and checkered history:
estimates of the gravitational wave luminosity have, over the last 30
years, ranged over more than four orders of
magnitude.\cite{mueller82a,finn89a,finn90b,finn91a,monchmeyer91a} It is
not simply the luminosity that is unknown: the waveforms themselves
are also entirely uncertain, leading to a further difficulty in
estimating the detectability of this source. (Examples in the
literature can be found in the citations\cite{saenz78a,saenz79a,saenz81a,finn89a,finn90b,finn91a,monchmeyer91a,zwerger98a}.)
Nevertheless, it is still possible to
evaluate what is required of stellar core collapse in order that it be
observable in a given detector.\cite{finn97h}

Suppose that the waveform from supernovae is given by
\begin{eqnarray}
h_{+} &=& {2{\rm M}_\odot\over r}\alpha f_{+} m(t)\\
h_{\times} &=& {2{\rm M}_\odot\over r}\beta f_{\times} m(t)
\end{eqnarray}
$\alpha$ and $\beta$ are constants, $f_{+}$ and $f_\times$ are
functions of the relative orientation of the source with respect to
the detector, and $m(t)$ is some function of time which we leave
undetermined for now. The power radiated into each polarization mode
is given by
\begin{eqnarray}
\dot{E}_{+} &=& \alpha^2\left<f^2_+\right>{\rm M}_\odot^2|\dot{m}|^2\\
\dot{E}_{\times} &=& \beta^2\left<f^2_{\times}\right>{\rm
M}_\odot^2|\dot{m}|^2,
\end{eqnarray}
where $<>$ signifies an {\em average\/} over a sphere centered on the
source.

Now assume that equal power is radiated into the two polarization
modes. Then we can write $\alpha$ and $\beta$ in terms of a single
parameter $\epsilon$ as
\begin{eqnarray}
\alpha^2 &=& {\epsilon\over2{\rm M}_\odot\left<f^2_+\right>
\int dt\,|\dot{m}|^2}\\
\beta^2 &=& {\epsilon\over2{\rm M}_\odot\left<f^2_{\times}\right>
\int dt\,|\dot{m}|^2}. 
\end{eqnarray}
In terms of $\epsilon$ the power radiated into the $+$ and $\times$
polarization states is thus
\begin{equation}
\dot{E}_+ = \dot{E}_\times = 
{\epsilon{\rm M}_\odot|\dot{m}|^2 \over 2\int dt |\dot{m}|^2}.
\end{equation}

Finally, return to consider the time dependence of the waveform
$m(t)$. Note that, by the Parseval Theorem,
\begin{equation}
\int dt\,|\dot{m}(t)|^2 = \int df\,(2\pi f)^2|\widetilde{m}(f)|^2,
\end{equation}
where $\widetilde{m}$ is the Fourier transform of $m$. Assume, as
suggested by numerical simulations, that the (real) power radiated per
unit bandwidth is approximately constant for frequencies in the
interval $(f_{\min}, f_{\max})$ and falls off rapidly
outside that band, with $f_{\min}$ on order 100~Hz and
$f_{\max}$ approximately 1~KHz. In this approximation,
\begin{equation}
\left(2\pi f|\widetilde{m}(f)|\right)^2 \simeq 
{\epsilon M_\odot\over f_{\max} - f_{\min}}.
\end{equation}

We can now evaluate the SNR we expect from core collapse supernova
gravitational waves incident on the detector.  Current calculations
suggest $\epsilon$ in the range $10^{-9}$--$10^{-8}{\rm M}_\odot$,
with peak power in the 200--300~Hz
band.\cite{finn91a,monchmeyer91a,burrows96a} For convenience here,
suppose that the detector noise power spectral density $S_h$ is
approximately constant over the bandwidth of the signal ({\em i.e.,}
from $f_{\min}$ to $f_{\max}$) and (optimistically) equal to its
approximate value at 100~Hz, and that $f_{\min}$ is very much less
than $f_{\max}$.  The mean-square signal-to-noise ratio is thus
\begin{eqnarray}
\overline{\rho^2} &\simeq& 1+
{\epsilon{\rm M}_\odot\over r^2}{1\over 2\pi^2S_{h}}
{1\over f_{\max}f_{\min}}\nonumber\\
&\simeq&
1+2.3{\epsilon\over10^{-8}}
\left({15\,{\rm Mpc}\over r}\right)^2
{10^{-46}\,{\rm Hz}^{-1}\over S_{h}} 
{100\,\mbox{Hz}\over f_{\min}}
{1\,\mbox{KHz}\over f_{\max}}
\end{eqnarray}
The distance of 15~Mpc is the range to the center of the Virgo Cluster
of galaxies: if supernovae can be reliably observed to distance, we
can expect a rate of approximately three per year. Reliable observation
of millisecond bursts at a rate of three per year, however, requires
an SNR $\overline{\rho^2}$ of somewhat more than 30.  Thus, without a
very optimistic efficiency $\epsilon$, we can't expect to be able to
observe supernovae much beyond our own galaxy, and certainly not out
to the Virgo cluster.

\subsection{Periodic Sources of Gravitational Radiation}\label{sec:periodic}

All the anticipated sources of periodic gravitational waves for the
ground-based detectors now under construction involve tapping the
stored rotational energy of rapidly rotating neutron stars.  This
means that the neutron star must be in some way non-asymmetric.  The
strength of the radiation depends on the degree of asymmetry.  All the
uncertainty associated with periodic sources of gravitational
radiation arises either with the mechanism for producing the asymmetry
or the degree of asymmetry.

At this writing four different kinds of asymmetries, or mechanisms for
producing asymmetries, are discussed as possibly leading to detectable
gravitational waves.  We discuss these in subsections
\ref{sec:nonaxis}, \ref{sec:precess}, \ref{sec:thermal},
\ref{sec:rmodes} below. Additionally, we discuss observational
constraints on the radiation from isolated pulsars in subsection
\ref{sec:observ}, and some issues related to the detection of periodic
radiation in subsection \ref{sec:detectPeriod}. 

\subsubsection{Non-axisymmetric rotators.}\label{sec:nonaxis}
All but the youngest neutron stars have a solid, crystalline crust,
covered by a fluid surface and with a fluid interior.  The fluid
surface and interior adjust themselves to the star's rotation,
remaining always axisymmetric and, therefore, not contributing to any
gravitational radiation.  As the star cools or spins-down (owing to,
{\em e.g.,} magnetic multipolar radiation if it is a pulsar) the shape
of its crust cannot adjust continuously to its new
conditions.  The stresses in the crust build until the crust
fractures, relieving the stress.  The final crust shape is likely to
be non-axisymmetric and responsible for gravitational radiation as the
star rotates.

Suppose that the star is rotating about a principal axis of its moment
of inertia tensor with rotational rate is $f$. Let $I_3$ be the 
moment of inertia along the axis or rotation, $I_1$ and $I_2$ be
the other two principal moments of inertia, and define
$\epsilon$ to be the difference between $I_1$ and $I_2$ relative to
$I_3$: 
\begin{eqnarray}
\epsilon &=& \left(I_2-I_1\right)/I_3.
\end{eqnarray}
Setting aside the very slow spin-down of the system as its angular
velocity changes and averaging over the angles that describe the
relative orientation of the pulsar with respect to the detector, the
characteristic radiation from this system is given by
\begin{eqnarray}
h(t) &\simeq& h_0\cos(4\pi f t+\phi_0)\label{eq:periodic}\\
\overline{h}_0 &=& {32\pi^{2}\sqrt{2}\over 5}{\epsilon f_0^{2} I_3\over r}\\
&=&4.8\times10^{-26}
{I\over 10^45\,\mbox{g cm}^{-3}}
{\epsilon\over10^{-6}}
{10\,\mbox{Kpc}\over r}
\left(f\over 300\,\mbox{Hz}\right)^3.
\end{eqnarray}

The power radiated gravitationally through this mechanism depends,
through $\epsilon$, on the degree of asymmetry that can be supported
by the neutron star crust.  Alpar and Pines\cite{alpar85a} have looked
at the structure of the crust and the likely strain that it can
support.  For our purposes it is instructive to look at the most
extreme possibility they considered: that the crust is well
approximated as a pure Coulomb-lattice crust.  Such a lattice could
sustain a strain some $10^3$ to $10^4$ times as much as is typical of
terrestrial material.  When the maximum allowable strain is supposed
to be supported by the solid part of the neutron star (which is only a
small fraction of the entire star), one arises with a maximum
$\epsilon$ of approximately $10^{-6}$.

This is an extreme value: it depends on the crust being a pure Coloumb
lattice, assumes that some mechanism has led it to be stressed to its
fracture point, and that the corresponding strain is principally
quadrupolar. For young neutron stars one might imagine this conspiracy
of circumstance possible; however, for older neutron stars plastic
flow of the crust would lead to relaxation over the age of the most
rapidly rotating neutron stars --- the so-called millisecond pulsars
---, reducing the maximum $\epsilon$ for these systems to no greater
than $4\times10^{-10}$.

\subsubsection{Observational constraints}\label{sec:observ}

There is observational evidence that, at least for the older,
millisecond pulsars, $\epsilon$ cannot be much greater than this limit.
The power $L$ radiated gravitationally by a spinning neutron star
comes directly from the star's rotation; consequently, radiation
back-reaction must slow the star in such a way that energy is
conserved. This leads to a slow spin-down of the star: if $P$ is the
spin period, then the rate of spin $\dot{P}$, assuming that
gravitational radiation reaction is the only source of spin-down, is
\begin{equation}
\dot{P} = -{LP^3\over\left(2\pi\right)^2I}.
\end{equation}
Since the radiated power $L$ is proportional to $\epsilon^2 I/P^6$
(cf.\ \ref{eq:power}) the measured period and period derivative place
a strict upper limit on $\epsilon$ for isolated pulsars.

The spin-down rate ($\dot{P}$) of most pulsars has been measured. If
we take the most extreme view and ascribe all of the spin-down to
angular momentum carried off by gravitational waves, the oblacity of
millisecond pulsars still can not exceed $10^{-9}$ for most
millisecond pulsars.\cite{alpar85a}

For a few young, isolated pulsars, timing observations are so good
that we can place still stronger limits on $\epsilon$: limits that
exclude the possibility that a significant part of the spin-down is
owing to radiation reaction.  For these pulsars, not only the rate of
the spin $\dot{P}$ but also its second derivative $\ddot{P}$ has been
measured.  Using only that the rotational energy of the star is
proportional to $I/P^2$ and that the radiated power is proportional to
$P^{-n}$ one can quickly show that
\begin{equation}
{P\ddot{P}\over\dot{P}^2} = 1-n.
\end{equation}
If the spin-down is due to quadrupole gravitational radiation
reaction, $n$ is equal to $5$ and higher-order radiative 
moments would lead to larger $n$. On the other hand, if the spin-down
is due to, say, magnetic dipole radiation (from the rotation of the
pulsar's magnetic dipole moment), $n$ is equal to $3$. There are no
isolated pulsars for which the measured $P\ddot{P}/P^2$ yields an $n$
approaching $5$. In particular, for the Crab pulsar, which is the
prototypical young pulsar, the measured $n$ is approximately $2.5$.
This is strongly inconsistent with the suggestion that gravitational
radiation damping plays an important --- let alone dominant --- role
in the spin-down of the Crab, or any other known, pulsar. Thus, we
must conclude that effective oblacity $\epsilon$ of pulsars is, in
fact, quite small and the radiation quite weak compared to the
detector noise.

\subsubsection{Detecting periodic sources}\label{sec:detectPeriod}

That the signal amplitude is small is, by itself, not of overwhelming
concern.  Consider for a moment the signal-to-noise associated with an
observation of a periodic signal, as is given in equation
\ref{eq:periodic}, over a period $T$ long compared to the periodicity
of the radiation $1/f_{\mbox{\small gw}}$. The SNR is given by 
\begin{equation}
\rho^2 = 1 + 4\int_0^\infty df{\left|\widetilde{h}(f)\right|^2\over
  S_h(f)}. \label{eq:snr1} 
\end{equation}
Since the signal is monochromatic over the observation period,
$\widetilde{h}$ is appreciable only at frequencies near $f_{\mbox{\small
    gw}}$; consequently, we can set $S_h(f)$ equal to
$S_h(f_{\mbox{\small gw}})$ in equation \ref{eq:snr1}. Then, invoking
The Parseval Theorem, we can rewrite equation \ref{eq:snr1} in the
time domain:
\begin{eqnarray}
  \rho^2 &=& 1+4\int_{0}^{\infty}df{\left|\widetilde{h}(f)\right|^2\over
    S_h(f)}\\
  &=&1+2\int_{-\infty}^{\infty} df{\widetilde{h}(f)^2\over
    S_h(f_{\mbox{\small gw}})}\nonumber\\
  &=&1+{2\over S_h(f_{\mbox{\small gw}})}\int_0^T
    dt\,h(t)^2\nonumber\\ 
  &=& 1+h^2_0{T\over S_h(f_{\mbox{\small gw}})};
\end{eqnarray}
{\em i.e.,} {\em for periodic sources the SNR grows with the
observation period.} As a result, a signal with a small peak amplitude
can be detected much more readily if it is periodic than if it is a
burst signal of finite duration.

In reality, there are practical limits to how large the SNR can be
made.  An observation can only be so long: for LIGO an observation
lasting a full year would be quite long.  Additionally, the signal is
not quite as simple as we have assumed.  The detector's motion with
respect to the source is non-uniform, owing both to Earth's rotation
about its axis and its orbit about the sun.  The time-dependent
Doppler shift of the signal leads to a frequency modulation, which
depends on the position of the source in the detector's sky.  In order
to obtain the growth of $\rho^{2}$ with time that we found above that
frequency modulation must be removed.  

Neither of these issues is significant if we know the position of the
source on the sky and its frequency. If, on the other hand, we are
contemplating a ``blind'' search, across the sky or over a wide
bandwidth in frequency, the story changes. If we are only interested
in one frequency, say $f_0$, then by folding the data we can find
$\widetilde{h}(f_0)$. If we are interested in a wide range of
frequencies, however, the most effective way to find
$\widetilde{h}(f)$ in that band involves a Fourier transform over the
interesting bandwidth. For a significant bandwidth (say, several
hundred Hz) and an observation lasting several months this is a
multi-gigapoint FFT. If, in addition, we are interested in searching
over the sky, then we must demodulate differently for different points
in the sky. For a several month observation, the number of independent
patches in the sky is quite large and the computational resources
required exceed any that might be available now or in the forseeable
future.\cite{brady98a} Thus, different ways of searching for periodic
sources must be developed, which will not have the same growth rate of
SNR with time.

Finally --- and this is true even if we are interested only in one
point on the sky and one frequency --- as the observation gets longer,
we must become concerned about the stability of our instrument and the
characterization of its noise.  One cannot expect the noise level to
remain stationary over indefinitely long periods; additionally, as we
observe for longer periods, we are, in an important sense, exploring
in greater and greater detail the character of the noise in very
narrow bandwidths.  As the observation period increases we must ask,
with an increasing degree of concern, how confident we are that there
are no technical noise sources, such as weak, drifting oscillators,
that may be masquerading as signal over the period of our observation.

\subsubsection{Precession.}\label{sec:precess}
An axisymmetric neutron star, rotating about it symmetry axis, does
not radiate gravitationally.  On the other hand, if the angular
momentum is not coincident with the symmetry axis --- {\em e.g.,} if
the star is precessing --- then it will radiate gravitationally. 
Misalignment of an axisymmetric neutron star's angular momentum and
body axes could arise as the result of crustal fractures associated
with a neutron star quake.

The same observational constraints that apply to gravitational
radiation arising from the rotation of a non-axisymmetric neutron star
about a principal axis also apply to radiation arising from precession
of an axisymmetric neutron star (cf.\ \S\ref{sec:observ})

While neutron star precession is --- in principle --- possible, if the
neutron star is also a pulsar, the precession should also manifest
itself as periodic variations in the electromagnetic pulse shape.  At
present there is no observational evidence for pulse shape variations
induced by free precession.  This may be because the misalignment is
too small to be observed in the pulse shape or because the stresses
associated with misalignment quickly bring the star back in to
alignment.

Gravitational radiation associated with precession of an axisymmetric
star occurs at both the rotational frequency and twice the rotational
frequency\cite{zimmerman79a}; consequently, it can be distinguished
from the radiation associated with a fully non-axisymmetric star
rotating about a principle axis. Interestingly, observing the
amplitude of the radiation at both the rotation frequency and twice
the rotation frequency allows one to determine the all the angles that
characterize the orientation of the star relative to the line-of-sight
(LOS): the angle between the angular momentum and the LOS as well as
the angle between the body axis and the angular momentum. If the star
is also observable as a pulsar, then one can test models of pulsar
beaming, since, together with the observed pulse shape, these make
predictions about the angle between the magnetic axis, the LOS and the
magnetic axis.

\subsubsection{Thermally driven non-axisymmetry.}\label{sec:thermal}
Timing of the
x-ray emission from several accreting neutron stars has revealed
quasi-periodic variability that can be explained as arising from the
rapid rotation of the underlying neutron star. An intriguing
coincidence in these observations is that the rotation rate of all
these systems appears to be close to equal. This suggests that there
is some underlying mechanism that insures that accretion spins these
stars up to --- but not beyond --- this limiting angular velocity. One
possibility is that the rotation rate is limited by gravitational
radiation reaction.

How might gravitational radiation limit the rotation rate of an
accreting system? If the accretion leads to a non-axisymmetry in the
neutron star then, as the star spins-up, the angular momentum radiated
by this rotating non-axisymmetry increases until it balances the
angular momentum accreted, limiting the stars rotation rate. The
angular momentum radiated is, like the radiated power, a strong
function of angular velocity ($\dot{J}$ is proportional to
$\Omega^{5}$, where $\Omega$ is the angular rotation rate); so, it is
not surprising that the limiting angular velocity should be similar
for these systems.

Proposals like this are characteristically made for systems where
there appears to be some upper limit to the rotation rate.  To be
plausible, there must be some universal mechanism whereby the same
process that spins the star up also leads to a non-axisymmetry that
can cause a radiative loss of angular momentum.  Recently Bildsten
\cite{bildsten98a} offered some promising ideas for a mechanism like
this that would operate in rapidly accreting, low magnetic field
neutron stars like Sco~X-1.

At the core of Bildsten's proposal is the observation that localized
heating of the neutron star owing to non-isotropic accretion leads to
differential electron capture rates in the neutron star fluid. These
lead, in turn, to density gradients as nuclear reactions in the
neutron star adjust its composition. Bildsten suggested that, if the
rotation axis is not aligned with the accretion axis and if some other
mechanism (in Bildsten's original suggestion, a magnetic field) can
break the symmetry still further, these density gradients may form in
a non-axisymmetric fashion.

There are two big ``ifs'' in this proposal.  Even if the accretion
disk is misaligned with the star's rotation axis the density
perturbations will be distributed symmetrically about the star's
rotation axis unless some other mechanism can be shown to break the
symmetry further.  Additionally, though not recognized in the original
proposal, buoyancy forces will lead the density perturbations to
re-distribute themselves in the star symmetrically about its rotation
axis, significantly suppressing the gravitational radiation.  For
these two reasons the initial excitement over the Bildsten proposal
has dampened.  It should not be extinguished, however: the recognition
that accretion can lead, through pyro-nuclear reactions, to density
perturbations that may radiate gravitational is certainly sound and
new.  With time will come greater understanding of where and how this
effect may arise in nature, and that greater understanding may yet
include a robust mechanism for producing significant gravitational
radiation from accreting neutron stars.

\subsubsection{R-mode instability.}\label{sec:rmodes}
Also in the last year, Andersson
\cite{andersson98a,andersson98b} discovered a new class of unstable
perturbative modes of rotating relativistic stars. In the absence of
gravitational radiation these modes are all stable; however,
gravitational radiation back-reaction on the modes causes them to
undergo exponential growth, feeding off of the rotational energy of
the star.

These particular modes are unusual in two different respects:
\begin{enumerate}
\item In the absence of viscosity they
  are unstable for {\em all\/} angular velocities;
\item At leading order the radiation is entirely ``magnetic'' in
  character: {\em i.e.,} the radiation couples to the momentum density
  distribution and its time evolution, not the mass density
  distribution and its time evolution.
\end{enumerate}

\begin{figure}
  \epsfxsize=0.8\columnwidth
  \begin{center}
    \leavevmode\epsffile{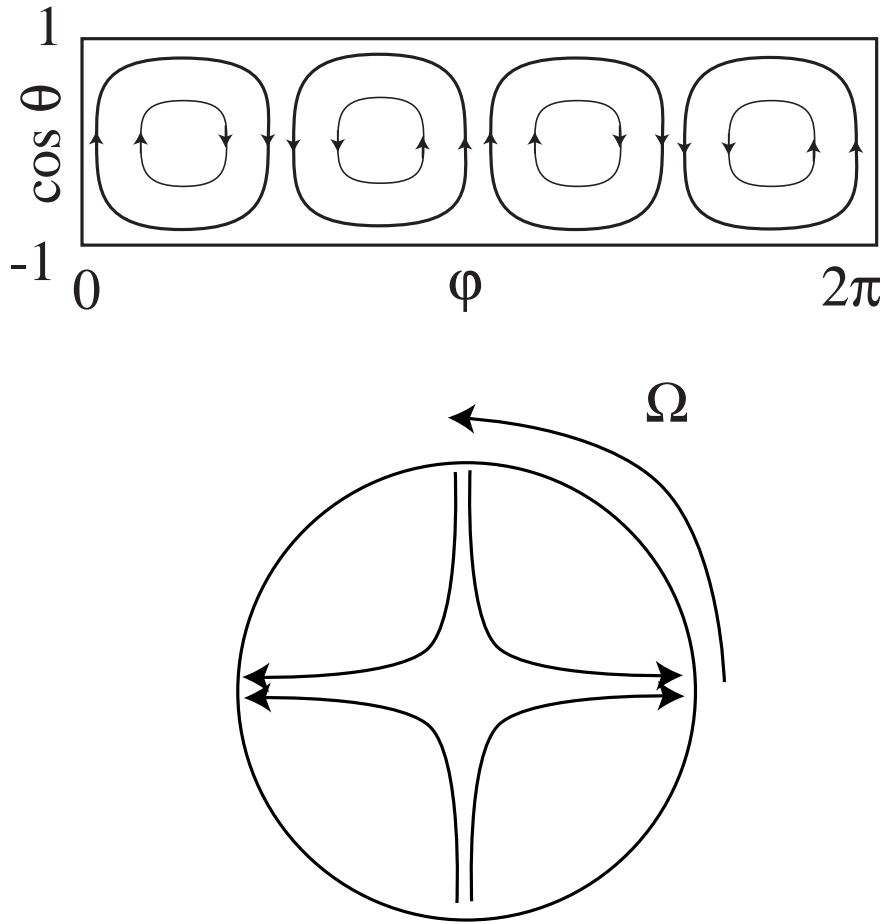}
  \end{center}
    \caption{Schematic illustration of the fluid flow-lines for the
    lowest order r-mode perturbation relative to the star's rotation. 
    The arrow indicate the relative direction of fluid flow on
    adjacent flow-lines.  In the upper panel the stars surface
    (co-latitude $\theta$, longitude $\phi$) is mapped to the plane;
    in the lower panel, the view is looking down along the stars
    rotation axis.  In both cases, the flow pattern is dragged along
    by the star's rotation.  At low rotation there is essentially no
    radial component to the fluids motion; so, we show only the
    angular components.}\label{fig:rmodes}
\end{figure}

Figure \ref{fig:rmodes} gives a schematic view of the character of the
fluid velocity perturbation, relative to the star's rotation, for the
lowest order radiating mode (magnetic quadrupole) and for low
rotational velocity.  The lines indicate fluid flow-lines; the arrow
indicate the relative direction of fluid flow on adjacent flow-lines.
At low rotation rates there is very little radial component to the fluid
motion; so, we show only the angular components.

In the absence of rotation, the fluid in the star can circulate in the
pattern shown with constant velocity: {\em i.e.,} there are no
restoring forces acting on the fluids inertia and the eigenfrequency
of the mode is zero. Correspondingly, the magnetic quadrupole moment
of this fluid mode has constant amplitude.  Rotation introduces
Coriolis forces, which act back on the fluid's inertia and cause the
circulation in these cells to be periodic. The quadrupole moment of
the fluid momentum now has a second time derivative; correspondingly,
it radiates gravitationally.

The power radiated gravitationally comes, ultimately, from the star's
rotational energy. The radiation thus carries off positive angular
momentum from the star. The greatest angular momentum can be carried
by the modes with azimuthal quantum number $m$ equal to $\pm\ell$; so,
focus attention on these modes. {\em Relative to the stars rotation,}
one of these modes is co-rotating and carries positive angular
momentum and one is counter-rotating and carries negative angular
momentum. It turns out, however, that both of these modes are dragged,
by the stars overall rotation, so that {\em viewed from an inertial
  frame both carry positive angular momentum.} The back-reaction of
the radiation adds negative angular momentum to each of these
modes; correspondingly, the co-rotating mode is damped and the
counter-rotating mode is anti-damped: {\em i.e.,} it grows. As it
grows, of course, it radiates more strongly, leading to greater
anti-damping: the mode undergoes exponential growth.

This general mechanism by which gravitational radiation reaction can
lead to amplification of a mode that is counter-rotating in the body
frame but co-rotating in the inertial frame was first discussed by
Zel'dovich in the context of rotating black holes. His concern was
that, through this mechanism, a rotating black hole should radiate ---
a result that anticipated the more general result of Hawking. The
first application to stars came from Chandrasekhar
\cite{chandrasekhar70a} and Friedman and Schutz\cite{friedman79a}
whose focus, however, was on a different set of modes.

Viscosity damps the R-modes. If the viscosity is large enough, then
the viscous damping exceeds the anti-damping caused by radiation
reaction and the mode is stabilized. Present understanding of the
viscosity of neutron star fluid suggests that there is a short
period in the life of a new-born neutron star, lasting perhaps one
year, when the mode may be unstable and a rapidly rotating star may be
a strong, nearly periodic source of gravitational radiation. The
radiation is ``nearly'' periodic because its amplitude is so great
that, over the course of the year, the stars angular rotation rate
may evolve from $10^3$ to $10^2$~Hz, simply due to the angular
momentum carried away in the radiation\cite{lindblom98a,owen98a}. 

\subsection{Stochastic Gravitational Radiation}\label{sec:stochastic} 

%\nocite{allen96a,allen97a,allen97c,christensen92a,flanagan93a,michelson87a,starobinskii79a,allen97b,ferrari98a,hils90a}

In the previous subsections we discussed burst and periodic sources of
gravitational radiation. In both cases the discussion focused on the
source and the radiation was characterized typically in terms of a
waveform $h(t)$, which depends on the details of a source
and its orientation with respect to the detector. 

The situation for a stochastic gravitational wave signal is different.
A stochastic gravitational wave signal is intrinsically random in
character. In particular, it is not generated by an isolated source,
it is not incident on the detector from a single direction, and it has
no characteristic waveform.

In fact, a stochastic signal can be treated as just another source of
detector noise. The stochastic radiation has an $h(t)$ that is
characterized solely by its correlation functions or associated
spectra; correspondingly, the detector output owing to the
stochastic signal is characterized in terms of correlation functions
or associated spectra.

Detection of any signal hinges on observing some characteristic
that distinguishes signal from noise. If a stochastic signal
appears in a detector no different than intrinsic detector noise, how
do we make the critical distinction that allows us to say we have
detected a signal? 

The essential difference between the action of a stochastic signal and
intrinsic detector noise is that stochastic radiation incident on two
detectors is correlated, and the correlation depends in a completely
predictable way on the relative orientation and separation of the
detectors.  Any gravitational wave signal --- stochastic or otherwise
--- can be resolved into a superposition of plane waves of different
frequencies and propagation directions bathing the detectors.  A
component of the radiation of given frequency and incident direction
drives two or more detectors coherently, with a phase delay that
depends on the incidence direction, detector separation and radiation
wavelength.  For components whose wavelength is much larger than the
separation between the detector, the phase difference is only weakly
dependent on the radiation wavelength or incident direction; so,
summed over incident direction there is a strong correlation between
the output of the detectors.  On the other hand, for radiation
components with wavelengths much smaller than the separation between
the detectors the phase difference depends strongly on the incident
direction and the wavelength; so, summed over incident direction, the
correlation between the output of the detectors is weak.  Thus, in the
presence of a stochastic gravitational wave background the output of
two or more detectors should show predictable correlations that depend
on their relative separations, relative orientations, and the
stochastic signal's spectrum\cite{flanagan93a}.

When considering sources of detectable stochastic gravitational
radiation for ground-based detectors, it is conventional to enumerate
the contributions of primordial origin: {\em e.g.,} radiation arising
during an inflationary epoch in the early universe\cite{kolb94a}, from
the decay of a cosmic string network\cite{vilenkin95a}, or from a
phase transition in the early
universe\cite{kamionkowski94a,kosowsky92a,kosowsky93a}.  Less
frequently discussed --- perhaps because it is so mundane --- are the
contributions arising from the {\em confusion limit\/} of discrete but
unresolved sources: {\em e.g.,} core-collapse
supernovae\cite{blair96a} or binary inspiral.  In fact, the
contribution at low-frequencies (periods of hours to minutes) to the
stochastic signal from unresolved galactic binary systems is expected
to be many times greater than the intrinsic detector noise of the
proposed space-based interferometric detector LISA
\cite{hils90a,folkner98a}.

Let's talk a bit more about how discrete, well-defined sources ---
whose waveforms may be well known --- can superpose to form a
stochastic signal of predictable spectrum. (We focus on burst sources
here; however, a similar story can be told for periodic
ones.\cite{hils90a})

Astrophysical sources of gravitational-wave bursts strong enough to be
observed by conceivable detectors are certainly independent events.
Setting aside the slow evolution of the source population on
cosmological timescales, the number of events whose radiation impinges
upon the receiver during any finite length observation is Poisson
distributed: if the event rate is $\dot{N}$ and the observation period
is $\tau$, then
\begin{equation}
P(n|\tau\dot{N}) = {\left(\tau\dot{N}\right)^n\over
  n!}e^{-\tau\dot{N}}.\label{eq:poisson}
\end{equation}

It is critical to recognize that the number of events exciting the
detector at any given moment is not the constant equal
to the product of the event rate and the signal duration.  The actual
number of events in the observation period varies.  Averaged over many
intervals it has a mean value, which is the product of the event rate
and the signal duration.  For example, table \ref{tbl:poisson} shows
the number of Poisson distributed events over consecutive five second
intervals, the rate calculated separately in each interval, and the
overall average rate.

\begin{table}
\begin{tabular}{rrrr}
Interval&\# events&``rate''&\\
0--5 s&3&0.6/s&number in\\
5--10 s&4&0.6/s&any interval\\
10--15 s&11&0.6/s&varies from\\
15--20 s&2&0.6/s&long-run rate\\
20--25 s&5&0.6/s&\\
\hline
0--25 s&25&1.0/s&long-run rate
\end{tabular}
\caption{Number of Poisson distributed events, with mean rate 1/s,
  occurring in consecutive five second intervals. Note that the number
  of events in any given interval varies: the rate is only the
  mean of the distribution of ``rates'' calculated over many
  intervals.}\label{tbl:poisson}
\end{table}

\begin{figure}
  \epsfxsize=0.8\columnwidth
\begin{center}
\leavevmode\epsffile{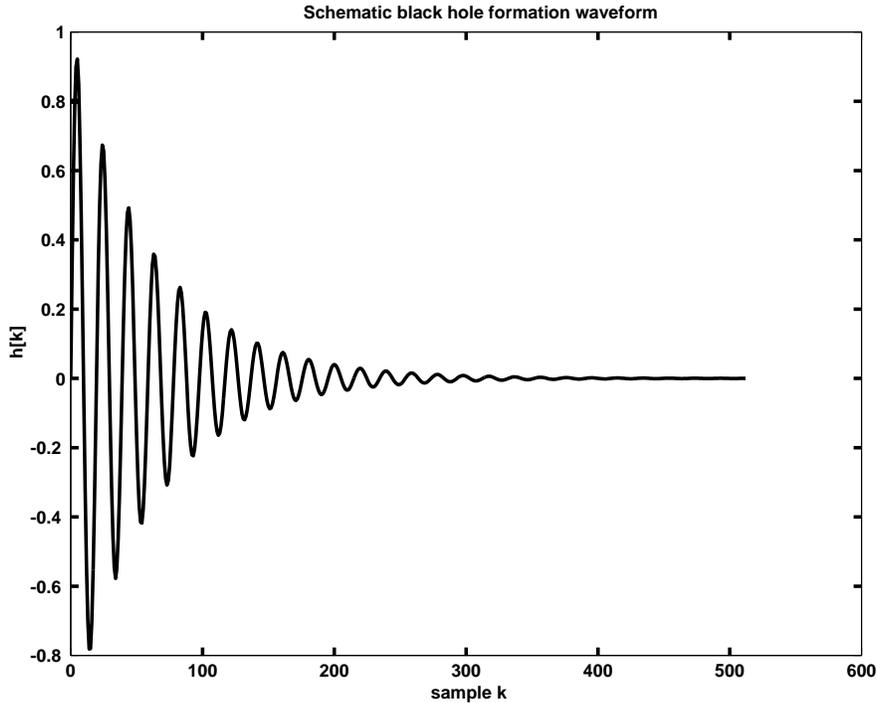}
\end{center}
  \caption{The prototypical waveform associated with an impulsive
    perturbation of a black hole: an exponentially damped
    sinusoid.}\label{fig:ringDown} 
\end{figure}

\begin{figure}
  \epsfxsize=0.8\columnwidth
\begin{center}
\leavevmode\epsffile{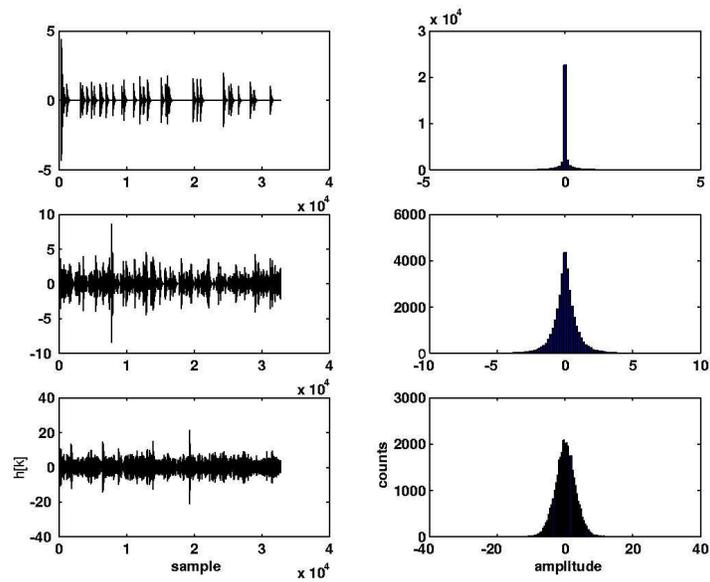}
\end{center}
  \caption{The left-hand panes show waveforms corresponding to the
    superposition of many discrete and idealized black hole formation
    events, with the number of events in a fixed interval Poisson
    distributed. The rate increases by an order of magnitde from the
    top to the middle panel, and again from the middle to the bottom
    panel. The right-hand panes show the distribution of wave
    amplitude derived from the corresponding left-hand
    panel.}\label{fig:ringDownHist}
\end{figure}

Now let's suppose that the ``events'' we are discussing are the
arrival at our detector of the initial wavefront associated with an
impulsive perturbation of a black hole. That waveform is a damped
sinusoid:
\begin{equation}
h(t) \propto e^{-2\pi f (t-t_0)/Q}\sin2\pi f(t-t_0) \qquad \mbox{for
  $t>t_0$.}
\label{eq:ringDown}
\end{equation}
The signal from a single, prototypical event is shown in figure
\ref{fig:ringDown}. The left-hand panels of figure
\ref{fig:ringDownHist}, on the other hand, 
shows $h(t)$ at the detector when signals exactly like these arrive at
the detector with different rates. In the top panel 
the signals are, for the most part, clearly
distinguishable. In the middle panel the rate is higher and separate
signals can be distinguished only in exceptional cases. In the bottom
panel the rate is higher still and the identity of the individual
signals is completely obscured.

Especially in the bottom panel of figure \ref{fig:ringDownHist}, the
signal $h(t)$ at any given $t$ is the superposition of a large, but
random, number of signals. Similarly, the amplitude of each
contributing signal is itself random (corresponding to the signal
amplitude at a random moment relative to the signal start time). In
circumstances like these we expect The Central Limit Theorem
\cite{mathews70a} to apply, leading to a normal distribution for
$h(t)$. On the other hand, at the lower rate pictured in the top
panel, there are not enough events superposed at any given moment for
us to expect $h(t)$ to exhibit a normal distribution. In the
right-hand panels of figure \ref{fig:ringDownHist} we show the
distribution of $h(t)$ taken from the left-hand panel and find that
these intuitions are borne out.

\begin{figure}
  \epsfxsize=0.8\columnwidth
\begin{center}
\leavevmode\epsffile{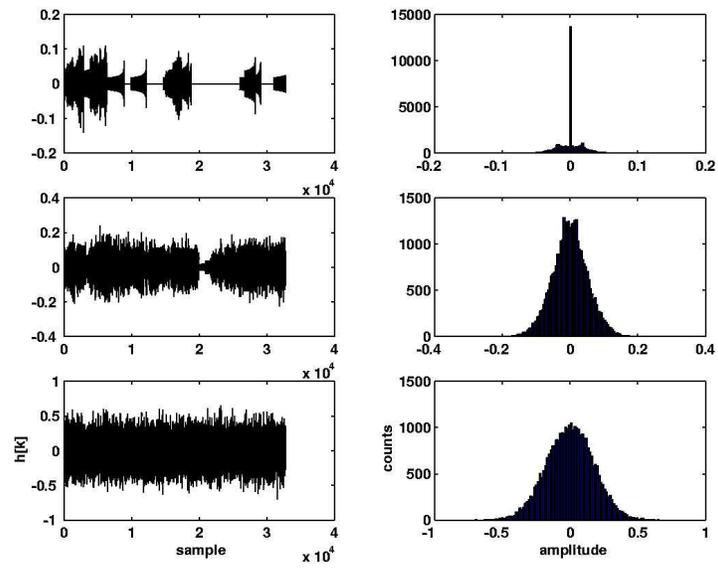}
\end{center}
  \caption{As in figure \protect{\ref{fig:ringDownHist}}, but with
    band-limited binary inspiral waveforms.}\label{fig:inspiralHist}
\end{figure}

Note that none of these conclusions depend in any way on the details
of the signal from an individual source: instead of the superposition
of damped sinusoids we could just as well have constructed $h(t)$ from
the superposition of binary inspiral signals in a fixed
bandwidth. We do exactly that in figure
\ref{fig:inspiralHist}. Note how the
distributions of $h(t)$ at high rate are, for superpositions of large
numbers of signals, identical ({\em i.e.,} they are normal
distributions). 

\begin{figure}
  \epsfxsize=0.8\columnwidth
\begin{center}
\leavevmode\epsffile{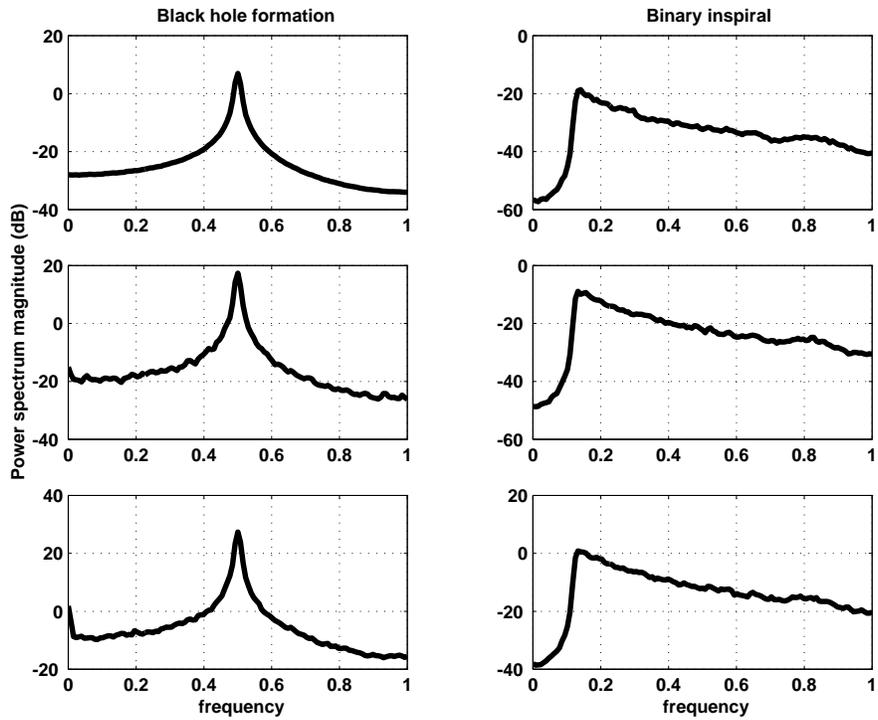}
\end{center}
\caption{Power spectra corresponding to the superpositions of black
  hole ringdown signals in figure \protect{\ref{fig:ringDownHist}} and
  binary inspiral signals in figure \protect{\ref{fig:inspiralHist}}.
  The power spectra reflects the underlying source of the stochastic
  signal (black hole ringdown or binary inspiral) while the amplitude
  is proportional to the rate of individual sources contributing to
  the signal.}
\label{fig:Pspectra}
\end{figure}

Despite the fact that the distribution of $h(t)$ arising from the
superposition of a large number of sources are identical, visual
inspection suggests that there are still differences. These
differences are associated with the correlations: the distribution of
the products $h(t)h(t+\tau)$ as a function of $\tau$. These are very
different for the damped sinusoids, which are characteristic of black
hole perturbations, and the ``chirps'', which are characteristic of
binary inspiral.  In figure \ref{fig:Pspectra} we show the power
spectral densities of the superpositions in figures
\ref{fig:ringDownHist} and \ref{fig:inspiralHist}. Note how the power
spectral densities of the stochastic signals formed from the random
superposition of events of a given character preserve the spectral
shape of the underlying signal, with the overall amplitude
proportional to the event rate. It is this property of the
superposition that permits us to predict the character of the
stochastic signal arising from the confusion limit of a large number
of sources; conversely, observation of a stochastic signal provides
us, through its amplitude and spectrum, information about the
underlying source, its rate and spatial distribution. 

This last point gives us the prospect of using a detected stochastic
background, arising from the superposition of unresolved sources, to
perform a source ``census'': a determination of the density of sources
in space.  This information, in turn, can shed light on astrophysical
process that are otherwise unobservable.  An excellent example of this
comes from LISA observations of the background from close white dwarf
binary systems (CWDBs).\cite{finn97e} These binary systems, which have
orbital periods ranging from days to hours, are so close that they are
optically unobservable as binaries. We know they exist, nevertheless,
because we can see their progenitor systems. CWDBs arise as one of the
end-points of binary star evolution, following a particularly
difficult to understand and model evolutionary stage where the two
stars are orbiting each other within a single envelope of gas. Once
they emerge from this final stage of {\em common envelope evolution\/}
they orbit each other as if point masses. Gravitational radiation
reaction leads to a slow, secular decay in the orbit of these systems
until they become so close that mutual tidal interactions lead to
rapid orbital decay or stellar disruption.

Let $dn/df$ be the number density of CWDBs with orbital frequency $f$.
The amplitude of the stochastic gravitational wave signal from this
population of objects is, at frequency $2f$, proportional to $dn/df$.
CWDBs are introduced into the population at a rate $d\dot{n}_{+}/df$
as they emerge from the final stage of common envelope evolution at
orbital frequency $f$; similarly, they leave the population at a rate
$d\dot{n}_{-}/df$ as they disrupt at orbital frequency $f$.  Knowing
the rate $d\dot{n}_{+}/df$ would provide us valuable information about
common envelope evolution, which we cannot obtain through optical
observations; similarly, the rate $d\dot{n}_{-}/df$ depends on the
white dwarf mass spectrum (we know the white dwarf equation of state
quite well) and knowing it would reveal the mass spectrum of binaries
emerging from the common envelope evolution stage.

We can determine both $d\dot{n}_{+}/df$ and $d\dot{n}_{-}/df$ from
LISA observation of the CWDB stochastic signal spectrum.  Observation
of this spectrum determines, as we have seen, $dn/df$.  Once injected
into the population, a CWDB's orbit evolves owing exclusively to
gravitational radiation reaction, which proceeds at the rate $df/dt$,
until it is removed from the population through coalescence or
disruption.  In steady state we thus have a {\em continuity
equation\/} governing the CWDB population:
\begin{eqnarray}
    {d\dot{n}_{+}\over df}-{d\dot{n}_{-}\over df} 
    &=& {d\over df}\left({dn\over df}{df\over dt}\right).
\end{eqnarray}
The orbital frequency evolution rate $df/dt$ is known (it is
proportional to $f^{11/3}$) and LISA observations will determine the
number density $dn/df$; consequently, from LISA observations we can
determine $d\dot{n}_{+}/df$ and $d\dot{n}_{-}/df$ and learn about the
end of common envelope evolution and the mass spectrum of white dwarfs
in CWDBs. 

\section{Conclusions}\label{sec:conclusions}

In these lectures I've tried to give a brief overview of how we think
about gravitational waves when we set out to detect them, and provide
a snapshot of current thinking on the anticipated wave sources.  Along
the way, I've tried to describe some of the science we can hope to do
once we can reliably detect gravitational wave sources.  

The principal difficulty in discussing the sources that we hope to
observe is our real lack of knowledge of their character.  As is often
the case, however, this difficulty is really a disguised opportunity:
when the detectors come on-line and we begin to detect gravitational
radiation sources, we will not simply be confirming what we already
know, but learning things entirely new about the cosmos!

%\bibliographystyle{unsrt}
%\bibliography{phyjabb,references,keys}

\end{document}